\documentclass[12pt]{article}

\usepackage{amsmath,amssymb,graphicx} %use drftcite in draftmode otherwise
\usepackage{epsf}
\usepackage{pstricks}
\usepackage{axodraw}
\usepackage{cite}

\newcommand{\beq}{\begin{eqnarray}}% can be used as {equation} or {eqnarray}
\newcommand{\eeq}{\end{eqnarray}}

\newcommand{\centeron}[2]{{\setbox0=\hbox{#1}\setbox1=\hbox{#2}\ifdim
                           \wd1>\wd0\kern.5\wd1\kern-.5\wd0\fi \copy0
                           \kern-.5\wd0\kern-.5\wd1\copy1\ifdim\wd0>\wd1
                           \kern.5\wd0\kern-.5\wd1\fi}}
\newcommand{\ltap}{\>\centeron{\raise.35ex\hbox{$<$}}
                   {\lower.65ex\hbox{$\sim$}}\>}
\newcommand{\gtap}{\>\centeron{\raise.35ex\hbox{$>$}}
                   {\lower.65ex\hbox{$\sim$}}\>}
\newcommand{\gsim}{\mathrel{\gtap}}

\newcommand\ZZ{\hbox{\zfont Z\kern-.4emZ}}
\font\zfont = cmss10 %scaled \magstep1
\newcommand{\sfrac}[2]{{\textstyle\frac{#1}{#2}}}
\newcommand{\oo}{0}
\def\tv#1{\vrule height #1pt depth 5pt width 0pt}
\def\tvbas#1{\vrule height 0pt depth #1pt width 0pt}
\newcommand{\sz}{\scriptstyle}

\begin{document}
\begin{titlepage}
\begin{flushright}
{\tt hep-ph/0210133} \\
Saclay t02/127\\
UCB-PTH-02/42\\
LBNL-51610
\end{flushright}

\vskip.5cm
\begin{center}
{\huge \bf Standard Model Higgs from}\\
\vskip.1cm
%{\huge \bf  from}\\
%\vskip.1cm
{\huge \bf Higher Dimensional Gauge Fields}
\vskip.2cm
\end{center}
\vskip0.2cm

\begin{center}
{\bf
{Csaba Cs\'aki}$^{a}$,
{Christophe Grojean}$^{b}$
{and Hitoshi Murayama}$^{c,d}$} \\

\end{center}
\vskip 8pt

\begin{center}
$^{a}$ {\it Newman Laboratory of Elementary Particle Physics\\
Cornell University, Ithaca, NY 14853, USA } \\
\vspace*{0.1cm}
$^{b}$ {\it Service de Physique Th\'eorique, CEA/DSM/SPhT CNRS/SPM/URA 2306,
CEA Saclay, F91191 Gif--sur--Yvette C\'edex, France} \\
\vspace*{0.1cm}
$^{c}$ {\it Department of Physics, University of California, Berkeley, CA
94720, USA} \\
\vspace*{0.1cm}
$^{d}$ {\it Theory Group, Lawrence Berkeley National Laboratory, Berkeley, CA
94720, USA} \\
\vspace*{0.3cm}{\tt  csaki@mail.lns.cornell.edu, grojean@spht.saclay.cea.fr,
murayama@hitoshi.berkeley.edu}
\end{center}

\vglue 0.3truecm

\begin{abstract}
\vskip 3pt
\noindent We consider the possibility that the standard model
Higgs fields may originate from extra components of higher dimensional
gauge fields.  Theories of this type considered before have had
problems accommodating the standard model fermion content and Yukawa
couplings different from the gauge coupling.  Considering orbifolds
based on abelian discrete groups we are lead to a 6 dimensional $G_2$
gauge theory compactified on $T^2/Z_4$.  This theory can naturally
produce the SM Higgs fields with the right quantum numbers while
predicting the value of  the weak mixing angle $\sin^2 \theta_{W} =
0.25$ at the tree-level, close to the
experimentally observed one. The quartic scalar coupling for the Higgs
is generated by the higher dimensional gauge interaction and predicts
the existence of a light Higgs.  We point out that one
can write a quadratically divergent
counter term for Higgs mass
localized to the orbifold fixed point.  However, we calculate these
operators and show that higher dimensional gauge interactions do not
generate them at least at one loop.  Fermions are introduced at orbifold
fixed points, making it easy to accommodate the standard model fermion
content.  Yukawa interactions are generated by Wilson lines.  They may
be generated by the exchange of massive bulk fermions, and the fermion
mass hierarchy can be obtained.  Around a TeV, the first KK
modes would appear as well as additional fermion modes localized at
the fixed point needed to cancel the quadratic divergences from the
Yukawa interactions.  The cutoff scale of the theory could be a few
times 10~TeV.
\end{abstract}

\end{titlepage}

\newpage

%\renewcommand{\thefootnote}{(\arabic{footnote})}

%%%%%%%%%%%%%%%%%%%%%%%%%%%%%%%%%%%%%%%%%%%%%%%%%%%%%%
%%%%%%%%%%%%%%%%%%%%%%%%%%%%%%%%%%%%%%%%%%%%%%%%%%%%%%
\section{Introduction}
\label{sec:intro}
\setcounter{equation}{0}
\setcounter{footnote}{0}
%%%%%%%%%%%%%%%%%%%%%%%%%%%%%%%%%%%%%%%%%%%%%%%%%%%%%%
%%%%%%%%%%%%%%%%%%%%%%%%%%%%%%%%%%%%%%%%%%%%%%%%%%%%%%

Theories with light elementary scalars seem unnatural, since
their masses receive quadratically divergent loop corrections, thus
one would expect their masses to be pushed up to the cutoff scale of
the theory.  This results in the well-known hierarchy problem of the
standard model (SM). The different approaches to solving the hierarchy
problem include eliminating the Higgs scalar entirely from the theory
(technicolor), lowering the cutoff scale (large extra dimensions and
Randall--Sundrum), or embed the Higgs field in a multiplet of a
symmetry group larger than the 4D Poincar\'e group (supersymmetry).

It has been observed quite a long time ago that besides supersymmetry
there may be other extensions of the 4D Poincar\'e group where scalars
could be embedded into, and thus perhaps protect their masses from
quadratic divergences.  The most natural such choice would be to use
the Poincar\'e group of a higher dimensional gauge theory, whereby
embedding elementary scalars into higher dimensional gauge multiplets.
In 1979 Manton~\cite{Manton} (and subsequently several
others~\cite{otherManton}) considered this possibility, where an extra
dimensional theory is compactified in the presence of a monopole in
the extra dimensions. The monopole background would then break higher
dimensional gauge invariance down to the SM group and result in a
negative mass square for some of the 4D scalars contained in the
higher dimensional gauge fields, thus resulting in successful
electroweak symmetry breaking. However it seems very difficult to
incorporate fermion matter fields into these theories. For a review of
such models see~\cite{Physicsreports}, for recent new ideas in this
direction see~\cite{Gia}. The idea of gauge symmetry breaking via
VEV's of scalars contained in the higher dimensional gauge fields was
further developed in the 80's by Hosotani~\cite{Hosotani}, and was
studied in detail in a string theory context~\cite{HosotaniString}
(for an early string realization of TeV size extra-dimensions see~\cite{Antoniadis90}).
For more recent work on the field theory side see
\cite{Hatanaka,ABQ,otherantoniadis,GIQ,HNS}. Four dimensional ``little
Higgs'' models motivated by this higher dimensional mechanism for
electroweak symmetry breaking were constructed and investigated
in~\cite{littlehiggs,minimallittle,otherlittle}.

Since our world is not supersymmetric, the key question in
supersymmetric extensions of the SM is to decide which operators are
softly breaking supersymmetry, that do not reintroduce quadratic
divergences and thus the hierarchy problem. Analogously, since we know
that our world does not have exact higher dimensional Poincar\'e
invariance, this symmetry needs to be broken, usually via
compactification of the extra dimensions.  Therefore an important task
is to understand in the context of these models what kind of
compactifications would maintain the absence of quadratic divergences,
and thus correspond to soft breaking of the symmetry. Clearly
compactification on tori would not reintroduce quadratic divergences,
however such compactifications are phenomenologically not so
interesting since they do not reduce the gauge group of the higher
dimensional theory, therefore one could only obtain scalar fields in
adjoint representations, which can not reproduce the SM. The next
simplest possibility is compactification on orbifolds~\cite{Orbifolds}, which we will
be considering in this paper. This enables one to reduce the size of
the unbroken gauge group by geometrically identifying regions in the
extra dimensional space, and thus allows one to obtain representations
other than adjoints under the unbroken gauge group to appear as 4D
scalars in the effective theory. Orbifold theories, with or without
Scherk--Schwarz compactification~\cite{SS}, have recently been used to
find a variety of interesting models of GUT~\cite{GUTbreaking,HMN} and
supersymmetry breaking~\cite{SSsusy,Susybreaking}.  5D theories
compactified on $S^1/Z_2$ do not naturally contain quartic couplings
for the scalars in the gauge fields. Therefore one is compelled to
look at 6D theories where the quartic scalar couplings are generated
by the higher dimensional gauge interactions. In the first part of the
paper we consider all possible models based on abelian 6D manifolds
using inner automorphisms which could lead to the SM as the low energy
effective theory. We identify the necessary compactifications for the
different choices of the gauge groups, find the resulting 4D scalars
that could serve as SM Higgs fields and calculate the prediction for
the weak mixing angle in the absence of brane induced gauge kinetic
terms. During this process we will identify a 6D gauge theory based on
the G$_2$ gauge group compactified on $T^2/Z_4$ (or generally on
$T^2/Z_k$ for $k\geq 4$) as the phenomenologically preferred choice in
theories of this sort. For this model we calculate the KK spectrum of
the orbifold theory and the quartic scalar coupling induced by the
gauge interactions.

However, an important part of the program is to check whether orbifold
compactifications reintroduce quadratic divergences or not.  In 5D
theories compactified on $S^1/Z_2$ there are no operators allowed by
gauge and Lorentz invariance~\cite{GIQ} that could reintroduce the
quadratic divergences.\footnote{The result concerns pure gauge theories
in the bulk. Once matter is introduced, in a supersymmetric context for instance,
some gauge invariant quadratic divergences can be generated at orbifold fixed points~\cite{NillesAndCo}.}
But in order to get the quartic scalar coupling
and also to be able to use a $Z_4$ orbifold projection we are
considering 6D theories.  In these models, we point out, there exist
operators which are a priori not forbidden by the Lorentz and gauge
symmetries of the orbifold theory, and thus could reintroduce the
quadratic divergence. These operators at the same time generate
tadpole terms in the effective action, thus the presence of such
operators would clearly be disastrous for a realistic model. Thus one
needs to investigate under what circumstances these tadpoles would be
generated.  This will be the focus of the second part of this paper.
We identify the potentially divergent brane induced operators for the
6D theories, and show that for $Z_2$ orbifolds parity invariance of
gauge interactions forbids the generation of this tadpole term.
However, for more complicated orbifolds parity invariance is broken by
the orbifolding, and thus we do not have a symmetry argument for the
absence of the tadpole terms.  Instead, we explicitly calculate the
one loop contribution for the tadpole term for the $T^2/Z_4$ theory
both based on an $SU(3)$ and a $G_2$ gauge group, and find that the
gauge contributions to the tadpole vanish.  It is then 
argued that even if generated at
higher loop order this term will not
destabilize the weak scale, due to the low value of the 6D cutoff scale.

Finally we consider adding fermions to the model at the orbifold fixed
points. Electroweak symmetry breaking is then triggered by the large
top Yukawa coupling. Fermions are introduced at orbifold fixed points.
Direct coupling of the fermions to the Higgs scalars would reintroduce
the quadratic divergences. However, one can generate fermion bilinear
interactions involving non-local Wilson-line operators which contain
the necessary Yukawa couplings for the leptons, by integrating out
vector-like bulk fermions which couple to the fermions localized at the
orbifold fixed points. For the quarks however one has to assume the
existence of these non-local operators in an effective theory approach
without being able to rely on an explicit mechanism to generate them.
We calculate the contribution of the Yukawa interactions to the Higgs
effective potential, and sketch the spectrum of the theory.

%%%%%%%%%%%%%%%%%%%%%%%%%%%%%%%%%%%%%%%%
%%%%%%%%%%%%%%%%%%%%%%%%%%%%%%%%%%%%%%%%
\section{The Choice of the Gauge Group}
\setcounter{equation}{0}
\setcounter{footnote}{0}
%%%%%%%%%%%%%%%%%%%%%%%%%%%%%%%%%%%%%%%

As discussed in the Introduction, we would like to find abelian
orbifolds of 6D gauge theories based on the gauge group $G$ which
could reproduce the bosonic sector of the SM without explicitly
introducing elementary scalars into the theory. We will restrict
ourselves to orbifolds using inner automorphisms, that is we use
elements $U$ of the group $G$ when doing the orbifold identifications.
Since we are using abelian subgroups of the original gauge group, the
rank of the gauge group will not be reduced~\cite{rank,Orbifolds}. Since we want
to obtain the $SU(2)\times U(1)$ electroweak gauge group after
orbifolding, the rank of $G$ has to be two. Thus there are only six
possibilities: $G= U(1)\times U(1), SU(2)\times U(1), SO(4)\sim
SU(2)\times SU(2), SU(3), SO(5)$ or $G_2$. The first two possibilities
are clearly unacceptable since $U(1)\times U(1)$ is not large enough
to accommodate the SM group, while $SU(2)\times U(1)$ can not produce
scalar fields that are not in the adjoint of $SU(2)\times U(1)$.  Thus
they can clearly not produce a SM Higgs doublet. The other four gauge
groups remain potential candidates, and we will consider them one by
one below.\footnote{There is another possibility that the rank of the
  unbroken gauge group is higher than two, while the unwanted part of
  the group breaks itself because of the anomaly~\cite{ABQ}.  In this
  case, one needs to rely on the Green--Schwarz mechanism for anomaly
  cancellation in the full theory.}

%%%%%%%%%%%%%%%%%%%
\subsection{G=SO(4)}

Since we are interested in abelian orbifolds using $SO(4) \simeq
(SU(2) \times SU(2))/Z_2$ group elements, the orbifold boundary
condition can only be in its maximal torus,\footnote{A maximal torus
  of the group is the maximal abelian subgroup generated
  by Cartan subalgebra, and is topologically a torus.}
\begin{equation}
   U = \left(
     \begin{array}{cccc}
       \cos \theta_1 & -\sin \theta_1 & 0 & 0\\
       \sin\theta_1 & \cos\theta_1 & 0 & 0\\
       0 & 0 & \cos\theta_2 & -\sin\theta_2 \\
       0 & 0 & \sin\theta_2 & \cos\theta_2
     \end{array} \right).
\end{equation}
For generic $\theta_{1,2}$, the group is broken as $SO(4) \rightarrow
SO(2) \times SO(2)$.  When $\theta_1=\theta_2$, the unbroken group is
enhanced to $(SU(2) \times U(1))/Z_2$.  However, the adjoint
representation of $SO(4)$ decomposes as ${\bf (3,1)+(1,3)} \rightarrow
{\bf 3_0}+{\bf 1_0} + {\bf 1_{+1}} + {\bf 1_{-1}}$ and hence there is
no candidate for Higgs doublet.  Finally when $\theta_1 = \theta_2 =
\pi$, the entire $SO(4)$ is unbroken.  None of these possibilities is
acceptable, thus we exclude the case $G=SO(4)$.

%%%%%%%%%%%%%%%%%%%%
\subsection{G=SO(5)}

The  maximal torus for $SO(5)$ is
\begin{equation}
   U = \left(
     \begin{array}{ccccc}
       \cos\theta_1 & -\sin\theta_1 & 0 & 0 & 0\\
       \sin\theta_1 & \cos\theta_1 & 0 & 0 & 0\\
       0 & 0 & \cos\theta_2 & -\sin\theta_2 & 0\\
       0 & 0 & \sin\theta_2 & \cos\theta_2 & 0\\
       0 & 0 & 0 & 0 & 1
     \end{array} \right).
\end{equation}
For generic $\theta_{1,2}$, the group is broken as $SO(5) \rightarrow
SO(2) \times SO(2)$.  When $\theta_2=0$, the unbroken group is
enhanced to $SO(3) \times SO(2) \simeq (SU(2)/Z_2) \times U(1)$.
However, the adjoint representation of $SO(5)$ decomposes as ${\bf 10}
\rightarrow {\bf 3_0}+{\bf 1_0} + {\bf 3_{\pm 1}}$ and hence there is
no candidate for Higgs doublet.  If we were to use the triplet anyway
for electroweak symmetry breaking, the $\rho$ parameter would not be
one at tree-level.

When $\theta_1=\theta_2$, the unbroken group is enhanced to $(SU(2)
\times U(1))/Z_2 \subset SO(4) \subset SO(5)$, a different embedding
of the electroweak group into $SO(5)$.  The adjoint representation
decomposes as ${\bf 10} \rightarrow {\bf 3}_0 + {\bf 1}_{\pm 1} + {\bf
  1}_0 + {\bf 2}_{\pm \frac{1}{2}}$, and hence we can obtain Higgs
doublets.  We however find $\sin^2 \theta_W =1/2$.  This would mean
that the dominant contribution to the gauge couplings would have to
come from the brane induced gauge kinetic terms, which is quite an
unnatural assumption.  Thus we exclude the case $G=SO(5)$.

%%%%%%%%%%%%%%%%%%%%
\subsection{G=SU(3)}

Breaking $SU(3)$ to $SU(2)\times U(1)$ can be achieved using any of
the $SU(3)$ group elements
\begin{equation}
        \label{twist}
U_k= {\rm diag} (\omega_k,\omega_k, \omega_k^{-2} ),
\end{equation}
where $\omega = e^{2\pi i/k}$, except for $k=3$ (in that case
$U_k\propto 1$ and the gauge group would remain $SU(3)$).  Since
$(U_k)^k=1$, we can have a $T^2/Z_k$ orbifold for any value of $k$
that would break $SU(3)\to SU(2)\times U(1)$. Let us now consider the
decomposition of the adjoint of $SU(3)$ under this breaking: ${\bf
  8}\to {\bf 3_0+1_0+2_3}$ where $2_3$ is a complex doublet, and the
$U(1)$ generator is ${\rm diag} (1,1,-2)=\sqrt{12} T_8$ in the $SU(3)$
fundamental.  If we want to redefine the normalization of the $U(1)$
generator so that the Higgs field has the standard 1/2 $U(1)_Y$
charge, we get that the low-energy gauge couplings would be related to
the $SU(3)$ coupling $g_3$ by $g=g_3$, $g'=\sqrt{3} g_3$, which would
result in $\sin^2 \theta_W =3/4$. This would mean that the dominant
contribution to the gauge couplings would again have to come from the
brane induced gauge kinetic terms, therefore we will not consider this
possibility either.

%%%%%%%%%%%%%%%%%%%%
\subsection{G=G$_2$}

This is the most interesting possibility. $G_2$ has two maximal
subgroups, $SU(3)$ and $SU(2) \times SU(2)$. The decomposition of the
$G_2$ adjoint under the $SU(3)$ subgroup is ${\bf 14}\to {\bf
  {8+3+\bar{3}}}$, where ${\bf 3+\bar{3}}$ form a complex~${\bf 3}$.
We can try to break the gauge group to the $SU(2)\times U(1)$
contained within the $SU(3)$ subgroup.  For this we can use group
elements that are contained within the $SU(3)$ subgroup, and use the
same $U_k$ elements as in (\ref{twist}). In order to find out what the
unbroken gauge group for the various choices of $k$ are, we need to
find out which generators remain invariant under the $Z_k$
transformation given by $U_k$.  For $k=2$ the $SU(2)\times U(1)$
subgroup of the $SU(3)$ subgroup of $G_2$ remains unbroken. However,
in the case of $k=2$ the orbifold action is $U_2= {\rm diag}
(-1,-1,1)$, which means that there are two additional generators from
the ${\bf 3+\bar{3}}$ which remain invariant, and the low-energy gauge
group will in fact be enlarged to $SU(2)\times SU(2)$ instead of the
desired $SU(2)\times U(1)$. Thus $k=2$ is excluded.  The case $k=3$ is
also excluded, since similarly to the $G=SU(3)$ case discussed above
the low-energy gauge group will be enlarged to $SU(3)$. However, for
any other value of $k\geq 4$ the low-energy gauge group is indeed
$SU(2)\times U(1)$. The value of the low-energy gauge couplings will
depend on which scalar will get an expectation value, since now there
could be two possibilities: the doublet that originates from the
adjoint of $SU(3)$ or the doublet from the ${\bf 3}$ of $SU(3)$.  We
have seen above that if it is the doublet from the $SU(3)$ adjoint
which is playing the role of the SM Higgs, we would get a prediction
for $\sin^2 \theta_W=3/4$, which is too far from the observed value.
The situation is however different, if the Higgs is contained in the
${\bf 3}$ of $SU(3)$. In this case the Higgs quantum numbers are given
by ${\bf 2_1}$, again with the $U(1)$ normalization ${\rm
  diag}(1,1,-2)$ in the $SU(3)$ fundamental.  Once we redefine the
normalization of $U(1)_Y$ so that this Higgs has the standard 1/2
charge we find that (see the Appendix for more details)
$g=g_{G_2}/\sqrt{2}$, $g'=g_{G_2}/\sqrt{6}$, and thus we find the
prediction for $\sin^2 \theta_W =1/4$, which is close to the observed
value. The difference can be made up either by small corrections from
brane induced kinetic terms or from running between the
compactification scale and the $Z$ mass scale.  This is similar to the
proposal of~\cite{DK}, and one needs to check the phenomenological
constraints on such models, as done in~\cite{CEKT}.  Note that the model also
contains the second scalar field coming from the $SU(3)$ adjoint. It
is an $SU(2)$ doublet but it has hypercharge $3/2$.  We will be
able to generate a positive quadratically divergent correction to its
mass square and thus this scalar will naturally decouple from the low
energy effective theory.

Thus we have found that the phenomenologically preferred models are
based on a 6D $G_2$ gauge theory, with a $Z_k$ orbifold for $k\geq 4$.
From now on we will concentrate on the simplest possibility with a
$Z_4$ orbifold.  Note that the $\sin^2 \theta_W$ values obtained for
the three possible gauge groups considered here are the same values
that Manton found~\cite{Manton} for the theories with monopole
backgrounds. This is not surprising, since these predictions are
purely based on group theory. Thus our conclusion is similar to
Manton's that the preferred models are based on the $G_2$ gauge group.

%%%%%%%%%%%%%%%%%%%%%%%%%%%%%%%%
\section{The KK spectrum of the 6D $G_2$ on $T^2/Z_4$}
\setcounter{equation}{0}
\setcounter{footnote}{0}
\label{sec:KK}
%%%%%%%%%%%%%%%%%%%%%%%%%%%%%%%%%

\subsection{KK decomposition and spectrum}

\begin{figure}
\centerline{\epsfxsize=.85\textwidth\epsfbox{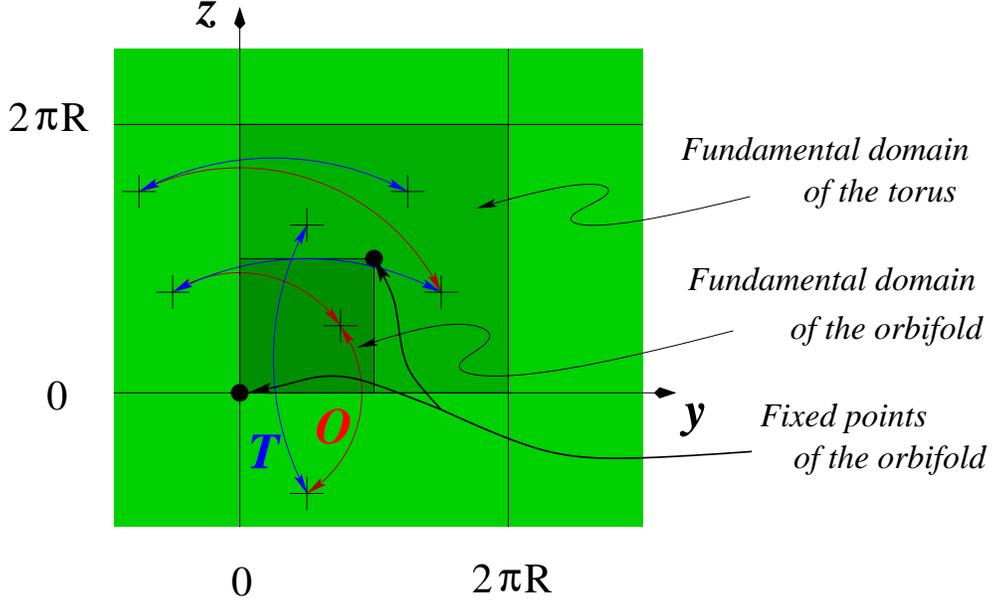}}
\caption{Symmetries of the orbifold. $T$ corresponds to torus
  identification while $O$ is the action of the orbifold symmetry. The
  fundamental domain of the orbifold space-time is the square $\pi
  R\times \pi R$ (however, for convenience, we will still normalize
  the KK modes by integration over the fundamental domain of the
  torus). Two points are left invariant by the orbifold action: the
  origin, $(0,0)$, and the point $(\pi R,\pi R)$. At these points, the
  $G_2$ gauge group of the bulk is broken to $SU(2)\times U(1)$ and
  gauge invariant potentially dangerous operators could be in
  principle generated}
\label{fig:orbifold}
\end{figure}

The decomposition of the $G_2$ fundamental under the $SU(3)$ subgroup
is ${\bf 7}\to {\bf {3+\bar{3}+1}}$.  A~useful basis for the
generators in the fundamental of $G_2$ is given in the
Appendix~\ref{App:algebra}.  The $Z_4$ orbifold symmetry acts on
space-time as a $\pi/2$ rotation on the extra dimensional coordinates,
as visualized on Fig.~\ref{fig:orbifold}.
The orbifold projection on gauge fields is defined by its action on
the fundamental representation:
\begin{equation}
  \phi (x,-z,y) = U \phi (x,y,z) \ \ \ {\rm with} \ \ \
  U={\rm diag}(i,i,-1,-i,-i,-1,1).
\end{equation}
The consistency of the orbifold projection with the gauge symmetry
dictates the transformation of the gauge fields:
\begin{eqnarray}
  A_\mu (x,-z,y)  & = &  U A_\mu (x,y,z) U^\dagger;
  \nonumber \\
  \label{eq:OrbifoldOnGauge}
  A_y (x,-z,y) & = & - U A_z (x,y,z) U^\dagger ;
  \\
  A_z (x,-z,y) & = & U A_y (x,y,z) U^\dagger .
  \nonumber
\end{eqnarray}
In our gauge basis, the action is written:
\begin{equation}
  \label{eq:AaM}
  A^a_M (x,-z,y) = {\tilde \Lambda}^{ab} {\tilde R}_{M}{}^{N} A^{b}_{N} (x,y,z)
\end{equation}
where ${\tilde \Lambda}$ and ${\tilde R}$ are $14\times 14$ and
$6\times 6$ non-diagonal matrices given for completeness in the
Appendix~{\ref{App:algebra}}.  To perform the Kaluza--Klein
decomposition, it is easier to first diagonalize the orbifold action,
which is achieved by defining light-cone like space-time coordinates
for the extra dimensions:
\begin{eqnarray}
  u^{\pm} = \sfrac{1}{\sqrt{2}} (y\pm i z)
  \ \ \
  {\it i.e.}
  \ \ \
  A^a_{\pm} = \sfrac{1}{\sqrt{2}} (A^a_y\mp i A^a_z).
\end{eqnarray}
Note that the metric is no longer diagonal and as a consequence, the
gauge propagator will for instance connect an index $+$ to a $-$ one:
\begin{equation}
  ds^2= dt^2 - dx_3^2 - 2 du^+ du^-
  \ \ {\it i.e.} \ \ g={\rm diag}
  \left(
    1,-1,-1,-1,\left( \begin{array}{cc}  &-1\\  -1 \end{array}\right)
  \right)
\end{equation}
We also need to redefine generators with well-defined hypercharge:
\begin{eqnarray}
  T^{1_\pm} = \sfrac{1}{\sqrt{2}} ( T^4 \pm i T^5)
  \ \ \
  & {\textit i.e. } &
  \ \ \
  A^{1_\pm}_M = \sfrac{1}{\sqrt{2}} (A^{4}_M    \mp i A^{5}_M)
  \nonumber \\
  T^{2_\pm}  = \sfrac{1}{\sqrt{2}} ( T^6 \pm i T^7)
  \ \ \
  & {\textit i.e.} &
  \ \ \
  A^{2_\pm}_M = \sfrac{1}{\sqrt{2}} (A^{6}_M    \mp i A^{7}_M)
\end{eqnarray}
Since the generators are non-hermitian, the metric in the gauge
indices is non-diagonal and as a consequence the gauge propagator will
for instance connect an index $1_+$ to a $1_-$ one or an index 9 to an
index 12:
\begin{equation}
  {\tilde g}^{ab} = 2 {\rm Tr} \left( T^a T^b \right)
  ={\rm diag}
  \left(
    1,1,1,
    \left( \begin{array}{cc}&1\\ 1 \end{array}\right),
    \left( \begin{array}{cc}&1\\ 1 \end{array}\right),
    1,
    \left( \begin{array}{cccccc}&&&1\\ &&&&1\\&&&&&1\\
        1\\&1\\&&1 \end{array}\right)
  \right)^{ab}
\end{equation}
As announced, in these systems of coordinates, the action of the
orbifold is diagonal and is simply (for short, we denote $u=(u^+,u^-)$
and $au=(a u^+,a^* u^-)$):
\begin{equation}
  A^a_M (x,iu) =
  {\Lambda}^{ab} {R}_{M}{}^{N} A^{b}_{N} (x,u)
\end{equation}
with
\begin{equation}
  R={\rm diag} (1,1,1,1,-i,i)
  \ \ \
  {\rm and}
  \ \ \
  \Lambda={\rm diag} (1,1,1,-i,i,-i,i,1,i,i,-1,-i,-i,-1)
\end{equation}
An eigenstate $\Phi$ associated to an eigenvalue $t$, {\it i.e.} satisfying:
\begin{equation}
  \Phi (x,iu) = t \, \Phi (x,u)
\end{equation}
can be written from an unconstrained field on the torus:
\begin{equation}
  \Phi (x,u) = \sfrac{1}{4} \left(
    \varphi (x,u)  + t^3 \varphi (x,iu) + t^2   \varphi (x,-u)
    +   t \varphi (x,-iu)
 \right)
\end{equation}
which leads to the KK decomposition: (we chose to normalize the
wavefunctions of the KK modes on the fundamental domain of the torus,
{\it i.e.}, the 4D effective action is obtained by integration of the
6D action over the fundamental domain of the torus)
\begin{eqnarray}
  \label{eq:KK}
  \Phi (x,u) =&&
  \sum_{p_y,p_z=1}^{\infty}
  \frac{ f_{p_y,p_z}(u) +t  f_{-p_z,p_y}(u) +
    t^2 f_{-p_y,-p_z}(u) + t^3 f_{p_y,-p_z}(u) }{4\pi R}\,
  \Phi^{(p_y,p_z)} (x)
  \nonumber \\
  +
  &&\sum_{p=1}^{\infty}
  \frac{ f_{p,0}(u) +t  f_{0,p}(u) +
    t^2 f_{-p,0}(u) + t^3 f_{p,0}(u) }{4\pi R}\,
  \Phi^{(p)} (x)
  + \frac{1}{2\pi R}\, \Phi^{(0)} (x)
\end{eqnarray}
where $f_{p_y,p_z}(u)$ are the KK wavefunctions on the square torus:
\begin{equation}
  f_{p_y,p_z}(u) =
  \exp{i(\sfrac{1}{\sqrt{2}R}(p_y-ip_z)u^+ 
    + \sfrac{1}{\sqrt{2}R}(p_y+ip_z)u^-)}
\end{equation}
Note that the last term in the KK decomposition~(\ref{eq:KK}) is
present only for an orbifold invariant field, {\it i.e.}, for $t=1$.
Indeed the other orbifold eigenvalues are not compatible with a flat
wavefunction, at least in the absence of a discontinuity
(discontinuities cannot be encountered for bosonic fields whose
equations of motion are of second order).

The KK modes $\Phi^{(p_y,p_z)} (x)$, $ \Phi^{(p)} (x)$ and $\Phi^{(0)}
(x)$ are canonically normalized in 4D and their masses are given by:
\begin{eqnarray}
  m^2_{(p_y,p_z)}=\frac{p_y^2+p_z^2}{R^2}, \ \ \ \ 
  m^2_{(p)}= \frac{p^2}{R^2}, \ \ \ \ 
  m^2_{(0)}=0
\end{eqnarray}
At the massless level, the spectrum contains $SU(2)\times U(1)$ gauge
bosons, $A_\mu^{1,2,3,8}$, as well as two complex scalar doublets: one
doublet coming from the $SU(3)$ adjoint representation,
$H=(A_-^{1_+},A_-^{2_+})=(A_+^{1_-},A_+^{2_-})^*$, which has
hypercharge $3/2$ in the normalization $Y={\rm diag} (1/2,1/2,-1)$ and
the other one coming from the $SU(3)$ fundamental and anti-fundamental
$h=(A_+^9,A_+^{10})=(A_-^{12},A_-^{13})^*$, with hypercharge $1/2$. In
order to get the preferred value of $\sin^2 \theta_W$ in the
low-energy theory, the SM Higgs should be identified with the
hypercharge $1/2$ field, while the other scalar should not get a VEV.
We will see that introducing fermions into the picture could naturally
achieve this breaking pattern.

%%%%%%%%%%%
\subsection{Higgs quartic coupling from 6D gauge interaction}

The 6D action contains a four gauge bosons interaction terms due to
the non-abelian nature of $G_2$ and, after compactification, the term
${\rm Tr} (F_{yz}F^{yz})$ gives rise to a quartic potential for the
Higgs scalars.  From the analysis above, it is easy to write the $A_y$
and $A_z$ gauge matrices in terms of the 4D canonically normalized
Higgs fields :
\begin{eqnarray}
  \hspace{-.1cm}
  A_y= \frac{1}{4\sqrt{2}\pi R}\left(
    \begin{array}{ccc|ccc|c}
      & & H_1 & & & \sqrt{\sfrac{1}{3}} h_2^* & \sqrt{\sfrac{2}{3}}h_1 \\
      & & H_2 & & & -\sqrt{\sfrac{1}{3}} h_1^* & \sqrt{\sfrac{2}{3}}h_2 \\
      \tvbas{10}
      H_1^* & H_2^* & & -\sqrt{\sfrac{1}{3}}h_2^* 
      & \sqrt{\sfrac{1}{3}}h_1^* &&\\
      \hline
      \tv{18}
      & & -\sqrt{\sfrac{1}{3}}h_2 & & &  -H_1^* & -\sqrt{\sfrac{2}{3}}h_1^* \\
      & & \sqrt{\sfrac{1}{3}}h_1 & & & -H_2^* & -\sqrt{\sfrac{2}{3}}h_2^* \\
      \tvbas{10}
      \sqrt{\sfrac{1}{3}}h_2 & -\sqrt{\sfrac{1}{3}}h_1 & & -H_1 & -H_2 &&\\
      \hline
      \tv{18}
      \sqrt{\sfrac{2}{3}}h_1^* & \sqrt{\sfrac{2}{3}}h_2^* & 
      & -\sqrt{\sfrac{2}{3}}h_1 & -\sqrt{\sfrac{2}{3}}h_2 &&
    \end{array}
  \right)&&
\label{Higgs1} \\
% \end{eqnarray}
% %
% \begin{eqnarray}
  A_z=\frac{-i}{4\sqrt{2}\pi R}\left(
    \begin{array}{ccc|ccc|c}
      & & H_1 & & & \sqrt{\sfrac{1}{3}} h_2^* & -\sqrt{\sfrac{2}{3}}h_1 \\
      & & H_2 & & & -\sqrt{\sfrac{1}{3}} h_1^* &- \sqrt{\sfrac{2}{3}}h_2 \\
      \tvbas{10}
      -H_1^* & -H_2^* & & -\sqrt{\sfrac{1}{3}}h_2^* 
      & \sqrt{\sfrac{1}{3}}h_1^* &&\\
      \hline
      \tv{18}
      & & \sqrt{\sfrac{1}{3}}h_2 & & &  H_1^* & -\sqrt{\sfrac{2}{3}}h_1^* \\
      & & -\sqrt{\sfrac{1}{3}}h_1 & & & H_2^* & -\sqrt{\sfrac{2}{3}}h_2^* \\
      \tvbas{10}
      -\sqrt{\sfrac{1}{3}}h_2 & \sqrt{\sfrac{1}{3}}h_1 & & -H_1 & -H_2 &&\\
      \hline
      \tv{18}
      \sqrt{\sfrac{2}{3}}h_1^* & \sqrt{\sfrac{2}{3}}h_2^* & 
      & \sqrt{\sfrac{2}{3}}h_1 & \sqrt{\sfrac{2}{3}}h_2 &&
    \end{array}
  \right)&&
  \label{Higgs2}
\end{eqnarray}
Here, $h$ ($H$) denotes the doublet scalar of hypercharge $1/2$
($3/2$).
After compactification to 4D, we obtain the following quartic
coupling:
\begin{equation}
  \label{eq:h4}
  V= \sfrac{1}{6} g^2_{4D} \left(
    |h|^4 + 3|H|^4 + 3(H^\dagger \sigma^a h) (h^\dagger \sigma^a H) -
    6 |h|^2 |H|^2 \right)
\end{equation}
where $\sigma^a, a=1\ldots 3$, are the Pauli matrices and
$g_{4D}=g_{6D}^{G_2}/(2\sqrt{2}\pi R)$ is the gauge coupling of the
low energy $SU(2)$ gauge group in 4D.

As we will see later, the doublet $H$ of hypercharge $3/2$ acquires a
quadratically divergent positive mass squared and decouples, while the
mass squared of the doublet $h$ of hypercharge $1/2$ can be protected
by a cancellation.  Therefore $h$ plays the role of the standard model
Higgs boson.  The quartic coupling $g^2_{4D}/6$ then predicts that the
tree-level Higgs mass is
$m_h^2 = \frac{1}{3} g^2_{4D} v^2 = M_Z^2=(91\ {\rm GeV})^2$.
This is similar to the situation in the MSSM, where the {\it
  maximal} tree-level value of the Higgs mass is $M_Z$. Loop
corrections to the quartic scalar coupling will modify this prediction
and push the Higgs mass to somewhat higher values.

%%%%%%%%%%%%%%%%%%%%%%%%%%%%%%%%%%%%%%%%%%
\section{Potentially Divergent Brane Induced Mass and Tadpole Operators}
\setcounter{equation}{0}
\setcounter{footnote}{0}

After identifying the interesting class of 6D models for electroweak
symmetry breaking, one needs to ask whether quadratically divergent
mass terms are indeed absent in this theory or not. The full higher
dimensional gauge group is operational in the bulk, and therefore one
does not expect quadratically divergent mass terms to be generated in
the bulk. However, the gauge invariance is reduced at the orbifold
fixed points, and one needs to find out if any brane localized
operators that would give quadratically divergent corrections to the
Higgs mass could be generated.

\subsection{General Discussion}

Gauge invariant operators are built using the field strength tensor
$F_{AB}$.  One could think that due to the reduced gauge invariance at
the orbifold fixed points one could use directly the 4D scalar
components of the gauge fields corresponding to the broken generators.
This is however not the case, as shown in \cite{GIQ}. The reason is
that the gauge transformation parameter $\xi^a$ has the same KK
expansion as the gauge fields themselves.  This means that while for
the broken generators $\xi^{\hat{a}}|_{fp}=0$, its derivatives with
respect to the extra dimensional coordinates do not vanish,
$\partial_i \xi^{\hat{a}}|_{fp} \neq 0$. Thus there is a residual
shift symmetry left from the higher dimensional gauge invariance even
for the broken generators, proportional to the derivative of the gauge
parameter, and one needs to consider invariants built from the field
strength tensor $F_{AB}(0)$ (in this section, the position ``0''
refers to the fixed point). Since $F_{AB}(0)$ transforms properly
under gauge transformations, its transformation law does not contain
any derivative pieces, and therefore it only transforms under the
unbroken gauge group as $F_{AB}(0)\to g(0) F_{AB}(0) g(0)^{-1}$ for a
finite gauge transformation $g(0)$ of the unbroken gauge group, since
the gauge transformation parameters for the broken generators vanish
at the fixed point.  The elements belonging to the broken part of the
group do not affect $F_{AB}(0)$ at the fixed point. The potentially
dangerous operators are linear in $F$, since their coefficient could
be quadratically divergent. Clearly, in 5D there is no such operator
allowed by Lorentz invariance, however in 6D the operator
\begin{equation}
  \label{badguy}
  {\rm Tr} ( U F_{yz}(0))
\end{equation}
is allowed, where $U$ is the group element used for the orbifold
projection.  This operator is clearly gauge invariant, since under
gauge transformations
\begin{equation}
  {\rm Tr} (U F_{yz}(0)) \to 
  {\rm Tr} (U g(0)  F_{yz}(0) g(0)^{-1}) ={\rm
  Tr} (U F_{yz}(0)),
\end{equation}
since $U$ commutes with the elements of the unbroken gauge group.
Similarly, any operator of the form ${\rm Tr} (U^n F_{yz}(0))$ for
$n=0,\ldots, k-1$ would also be allowed, but as we will see below
these all lead to the same set of allowed operators on the fixed
point.

For the case of $SU(3)\to SU(2)\times U(1)$ and $G_2\to SU(2)\times
U(1)$ using the orbifold based on $U={\rm diag}(i,i,-1)$ this operator
will be proportional to $F^8_{yz}(0)$, where the $8$ index refers to
the unbroken $U(1)$ generator within $SU(3)$. This term would contain
a {\it tadpole} for the scalar components of $A^8$, and through the
$f^{bc}{}_8 A^b_y A^c_z$ term contained within the field strength also
mass terms for the scalars that are supposed to play the role of the
SM Higgs. Therefore it is essential to find out under what
circumstances these operators are generated.

Generically, the operator in (\ref{badguy}) will pick out the field
strength tensor corresponding to the unbroken $U(1)$
components.\footnote{We thank M.  Quir\'os for this remark.} This can
be seen by examining the group matrix structure of $ {\rm Tr} (U
F_{yz}) = {\rm Tr} (U T^a) F^a_{yz}$. The orbifold projection is
telling us that $U T^a U^\dagger = t^a T^a$, where $t$ is the $Z_k$
parity of the particular generator.  From this $(1-t^a){\rm Tr} (U
T^a)=0$, thus (\ref{badguy}) can only contain elements from the
unbroken group. However, all elements of the unbroken Lie algebra can
be written as commutators of other Lie algebra elements in the
unbroken group, unless it is corresponding to a $U(1)$
factor.\footnote{Mathematically, we are saying that the derived
  algebra of a semi-simple algebra is the algebra itself.}  Since $U$
commutes with the unbroken generators the contributions to
(\ref{badguy}) vanish for all elements in the non-abelian component of
the unbroken part of the gauge group, and only the unbroken $U(1)$
factors can contribute.

Next we show that in the case of a $Z_2$ orbifold parity invariance
forbids the generation of (\ref{badguy}), however for $Z_4$ or other
higher $Z_k$ there is no discrete symmetry to forbid this operator.

\subsection{$Z_2$ orbifold}

First we consider a $T^2/Z_2$ orbifold.  The orbifold boundary
condition under $Z_2$ ($y \rightarrow -y$, $z \rightarrow -z$) for a
bulk scalar $\phi(x^\mu, y, z)$ is
\begin{equation}
  \phi(x, -y, -z) = U \phi(x, y, z).
  \label{eq:Z2phi}
\end{equation}
$U$ is an element of the gauge group that satisfies $U^2 = 1$.  We
want to show that there is parity invariance in the
Yang--Mills--scalar theory.  Obviously the Yang--Mills--scalar theory
on two-dimensional torus is parity invariant.  Therefore the only
condition to check is if the orbifold boundary condition is consistent
with parity.  In even-dimensional space, parity is defined by flipping
only one (or an odd number of) spatial coordinate.  Let us consider $y
\rightarrow -y$, $z \rightarrow z$.  Under this parity, the l.h.s. of
Eq.~(\ref{eq:Z2phi}) becomes $\phi(y, -z)$, while the r.h.s.
$U\phi(-y,z)$.  Because Eq.~(\ref{eq:Z2phi}) must hold for any $y$ and
$z$, $\phi(y,-z) = U \phi(-y, z)$, and the parity-transformed
Eq.~(\ref{eq:Z2phi}) holds.  In other words, the condition
Eq.~(\ref{eq:Z2phi}) is parity invariant.

The orbifold boundary condition for the gauge field is obtained
from the requirement that the covariant derivative of the bulk scalar
transforms covariantly under the orbifold boundary condition:
\begin{eqnarray}
  A_\mu (x,-y,-z)  & = &  U A_\mu (x,y,z) U^\dagger;
  \nonumber \\
  \label{eq:Z2A}
  A_y (x,-y,-z) & = & - U A_y (x,y,z) U^\dagger ;
  \\
  A_z (x,-y,-z) & = & - U A_z (x,y,z) U^\dagger .
  \nonumber
\end{eqnarray}

Now we try to identify the parity transformation of the gauge field
that preserves Eq.~(\ref{eq:Z2A}).  The gauge field transforms under
parity normally as
\begin{eqnarray}
  A_y (y,z) &\rightarrow& -A_y (-y, z), \nonumber \\
  A_z (y,z) &\rightarrow& A_z(-y, z).
\end{eqnarray}
Again, it is easy to see that the parity preserves the orbifold
boundary condition.  Under the parity, the operator of our concern
${\rm Tr} (U F_{yz})$ transforms to $-{\rm Tr} (U F_{yz})$.
Therefore, parity invariant interactions would no induce this term.

If there are bulk fermions present, one needs to check if parity
invariance is broken or not. Clearly for vector-like fermions one can
extend the definition of parity in the usual way, and we expect that
(\ref{badguy}) would not be generated. For more complicated
representations one would have to perform an explicit calculation to
check for the presence of this term.

\subsection{$Z_4$ orbifold}

Next we consider a $T^2/Z_4$ orbifold.  The orbifold boundary
condition under $Z_4$ ($y \rightarrow -z$, $z \rightarrow y$) for a
bulk scalar $\phi(x^\mu, y, z)$ is
\begin{equation}
  \phi(-z, y) = U \phi(y, z).
  \label{eq:Z4phi}
\end{equation}
$U$ is an element of the gauge group that satisfies $U^4 = 1$.  We
will now show that one can not define a parity invariance in this
theory.  Obviously the Yang--Mills--scalar theory on two-dimensional
torus is parity invariant.  Therefore the only condition to check is
if the orbifold boundary condition is consistent with parity.  Let us
consider $y \rightarrow -y$, $z \rightarrow z$.  Under this parity,
the l.h.s. of Eq.~(\ref{eq:Z4phi}) becomes $\phi(z, y)$, while the
r.h.s. $U\phi(-y,z)$.  Because Eq.~(\ref{eq:Z4phi}) must hold for any
$y$ and $z$, $\phi(z, y) = U^{-1} \phi(-y, z)$, and the
parity-transformed equation reads
\begin{equation}
   U^{-1} \phi(-y, z) = U \phi(-y, z).
\end{equation}
This equation is inconsistent unless $U^2=1$.  Therefore, the naive
definition of parity is not a symmetry of the theory.

One modification allows for a similar symmetry, which actually is a CP
rather than~P.  At the same time of flipping the sign of $y$, we take
complex conjugate of $\phi$, namely $\phi(y,z) \rightarrow
\phi^*(-y,z)$.  Then Eq.~(\ref{eq:Z4phi}) becomes
\begin{equation}
  \phi^*(z, y) = U \phi^*(-y, z).
\end{equation}
Using the same line of reasoning as before, the l.h.s. is rewritten as
\begin{equation}
  U^{-1*} \phi^*(-y, z) = U \phi^*(-y, z).
\end{equation}
This equation is consistent if $U^* U = 1$.  Indeed in the ``unitary
gauge'' where we treat $U$ as a fixed gauge element, we can always
diagonalize $U$ to be a pure phase matrix.  Then this condition is
indeed satisfied.  Therefore, the CP symmetry is still intact.

Under this CP, the bulk fields are complex conjugated, and
correspondingly, the gauge fields must be brought into the conjugate
representation $T^a \rightarrow - {T^{a}}^t$.  Then the transformation
properties of the gauge fields are:
\begin{eqnarray}
  A_y (y,z) &\rightarrow& A_y^t (-y,z) \nonumber \\
  A_z (y,z) &\rightarrow& -A_z^t (-y, z).
  \label{eq:Z4A}
\end{eqnarray}

Under the CP, the operator of our concern ${\rm Tr} (U F_{yz})$
transforms to $+{\rm Tr} (U F_{yz}^t)$.  Because the trace is
transpose-invariant, it is ${\rm Tr} (U^T F_{yz}) = {\rm Tr} (U
F_{yz})$.  Therefore the operator is CP invariant and hence CP does
not forbid its generation from loops.

However, there is another modification of parity that may be preserved
by the orbifold boundary condition.  Instead of the naive parity
$\phi(y,z) \rightarrow \phi(-y,z)$, we allow a gauge transformation on
top of it, $\phi(y,z) \rightarrow P \phi(-y,z)$ for $P \in G$.  Under
this parity, the l.h.s. of Eq.~(\ref{eq:Z4phi}) becomes $P \phi(z,
y)$, while the r.h.s. $U P\phi(-y,z)$.  Because Eq.~(\ref{eq:Z4phi})
must hold for any $y$ and $z$, $\phi(z, y) = U^{-1} \phi(-y, z)$, and
the parity-transformed equation reads
\begin{equation}
  P  U^{-1} \phi(-y, z) = U P \phi(-y, z).
\end{equation}
This equation is consistent if $P U^{-1} = U P$.  The question is if
you can find such $P$ within the gauge group. We will show in Appendix
\ref{App:G2element} that one can indeed find a group element that
satisfies this constraint for the case when $G_2$ is broken to
$SU(2)\times U(1)$ by the $Z_4$ orbifold.  However, this modified
parity still does not forbid the tadpole in (\ref{badguy}).  Under
this parity,
\begin{equation}
  {\rm Tr} (U F_{yz}) \rightarrow {\rm Tr} (U P (-F_{yz}) P^{-1})
  = - {\rm Tr} (U^* F_{yz}).
\end{equation}
Therefore, the allowed combination is
\begin{equation}
  i \left( {\rm Tr} (U F_{yz}) - {\rm Tr} (U^* F_{yz}) \right)
\end{equation}
while the sum is forbidden. This is still not enough to forbid the
mass term for the Higgs component.

%%%%%%%%%%%%%%%%%%%%%%%%%%%%%%%%%%%%%%%%%%%%%%%%%
\section{Tadpole Cancellation for $Z_4$ Orbifolds}
\setcounter{equation}{0}
\setcounter{footnote}{0}

We have seen above, that for $Z_2$ orbifolds parity forbids the
generation of the tadpole (\ref{badguy}), however for higher $Z_k$ (as
is needed for the $G_2$ model) we could not identify such a symmetry.
Therefore we need to explicitly calculate the coefficient of the
tadpole term to see whether or not it is indeed generated. For this,
we need to find the propagators on a $Z_4$ orbifold spacetime, which
can be done by generalizing the work of Georgi, Grant and
Hailu~\cite{GGH}.

%%%%%%%%%%%%
\subsection{Propagators for $Z_4$}

The orbifold constraints on the gauge fields, $A_M^a$, can be
implemented in terms of a set of unconstrained gauge fields on the
torus, $\mathcal{A}_M^a$:
\begin{eqnarray}
  &&A_M^a (x,u) = \sfrac{1}{4}
  (\mathcal{A}_M^a (x,u) +
  (R^{-1})_{M}{}^{M'} (\Lambda^{-1})^{aa'} \mathcal{A}_{M'}^{a'} (x,iu)
  \nonumber \\
  && \hphantom{spa} +
  (R^{-2})_{M}{}^{M'} (\Lambda^{-2})^{aa'} \mathcal{A}_{M'}^{a'} (x,-u) +
  (R^{-3})_{M}{}^{M'} (\Lambda^{-3})^{aa'} \mathcal{A}_{M'}^{a'} (x,-i u) )
\end{eqnarray}
In order to avoid further mixing of the fields we will choose to work
in the Feynman gauge $\xi=1$. The propagator for the unconstrained
fields $\mathcal{A}$ takes its usual expression in this gauge (we
denote by $\delta^{(2)}_{p-q}$ the product $\delta_{p_+-q_+}
\delta_{p_--q_-}$):
\begin{eqnarray}
  \langle \mathcal{A}^a_M (p) {\bar {\mathcal{A}}}^b_N (q) \rangle
  = G_{MN}^{ab} (p) \, \delta^{(2)}_{p-q}
  = - i \frac{{\tilde g}^{ab} g_{MN} }{p^2 - 2 p_+ p_-} \,
  \delta^{(2)}_{p-q}
\end{eqnarray}
where $g_{MN}$ and ${\tilde g}_{ab}$ are the space-time metric and the
gauge metric defined in Section \ref{sec:KK}.  Using the unitarity of
the matrices $R$ and $\Lambda$ and the fact that the unconstrained
propagator satisfies ($p_\pm=(p_y\mp p_z)/\sqrt{2}$, $p=(p_+,p_-)$ and
$a p=(a p_+,a^* p_-)$)
\begin{equation}
  (R^{-1})_{M}{}^{M'} (\Lambda^{-1})^{aa'}
  G_{M'N}^{a'b} (p_\mu,-i p)
  = G_{MN'}^{ab'} (p_\mu,p)
  R_{N}{}^{N'} \Lambda^{bb'}
\end{equation}
we obtain the gauge propagator on the $Z_4$ orbifolded torus:
\begin{eqnarray}
  \langle A^a_M (p) {\bar A}^b_N (q) \rangle
  &=&  \frac{G_{MN'}^{ab'} (p)}{4}
  (\delta_N^{N'} \delta^{bb'} \delta^{(2)}_{p-q}
  +
  R_{N}{}^{N'} \Lambda^{bb'} \delta^{(2)}_{p+iq}
  \nonumber \\
  &+& 
  R^2_{N}{}^{N'} {\Lambda^2}^{bb'} \delta^{(2)}_{p+q}
  + R^3_{N}{}^{N'} {\Lambda^3}^{bb'} \delta^{(2)}_{p-iq}
  )
  \hspace{.2cm}
\end{eqnarray}
In the same way, from the unconstrained ghost propagator:
\begin{eqnarray}
  \langle \mathcal{C}^a (p) {\bar {\mathcal{C}}}^b (q) \rangle
  = G^{ab} (p)\, \delta^{(2)}_{p-q}
  =  i \frac{{\tilde g}^{ab}}{p^2-2 p_+ p_-}
  \, \delta^{(2)}_{p-q}
\end{eqnarray}
we obtain the ghost propagator on the orbifold torus:
\begin{eqnarray}
  \langle C^a (p) {\bar C}^b (q) \rangle
  = \frac{G^{ab} (p)}{4} \left(
    \delta^{bb'} \delta^{(2)}_{p-q}
    + \Lambda^{bb'} \delta^{(2)}_{p+iq}
                                %\nonumber \\
    + {\Lambda^2}^{bb'} \delta^{(2)}_{p+q}
    + {\Lambda^3}^{bb'} \delta^{(2)}_{p-iq}
  \right)
\end{eqnarray}
Note that these relations can be easily generalized now to a general
$Z_k$ orbifold of $T^2$. The only difference will be that the
propagators will in general contain $k$ terms, with the $n^{th}$ term
containing $\Lambda^n$, and the matrix $R$ will be replaced by $R={\rm
  diag} (1,1,1,1,e^{-2\pi i/k},e^{2\pi i/k})$. The momentum conserving
delta functions on the $n^{th}$ term in the propagator is obtained by
expanding $\delta^{(2)}_{p-(R^n q)}$, where $n=0,1,\ldots ,k-1$.

\subsection{Explicit calculation of the tadpoles for $G_2$ and $SU(3)$
on $T^2/Z_4$}

\begin{figure}
\centerline{
\begin{picture}(250,110)%(110,110)(-20,0)
%%%%%%%%%%%%%%%%%%%%%%%%%%%%%%%%%%%%%%%%%%%%%5
  \Photon(-20,50)(40,50){3}{5} \Vertex(40,50){3}
  \Text(-40,50)[b]{$a, \pm$}
  \LongArrow(10,40)(30,40)
  \Text(10,30)[l]{$p$}
  \PhotonArc(70,50)(30,0,360){3}{20}
  \LongArrowArcn(70,50)(40,155,110)
  \Text(32,82)[l]{q}
  \LongArrowArcn(70,50)(40,250,205)
  \Text(37,15)[l]{r}
  \Text(60,-20)[c]{(a)}
%%%%%%%%%%%%%%%%%%%%%%%%%%%%%%%%%%%%%%%%%%%%%%%%%%%%%%%%
\end{picture}
\begin{picture}(50,110)%(110,110)(-20,0)
%%%%%%%%%%%%%%%%%%%%%%%%%%%%%%%%%%%%%%%%%%%%%5
  \Photon(-20,50)(40,50){3}{5} \Vertex(40,50){3}
  \Text(-40,50)[b]{$a, \pm$}
  \LongArrow(10,40)(30,40)
  \Text(10,30)[l]{$p$}
  \DashArrowArcn(70,50)(30,179,180){3}
  \LongArrowArcn(70,50)(40,155,110)
  \Text(32,82)[l]{q}
  \LongArrowArcn(70,50)(40,250,205)
  \Text(37,15)[l]{r}
  \Text(40,-20)[c]{(b)}
%%%%%%%%%%%%%%%%%%%%%%%%%%%%%%%%%%%%%%%%%%%%%%%%%%%%%%%%
\end{picture}
}
\vspace{.8cm}
\caption{\label{fig:tadpole}The gauge and ghost contributions to the tadpole
  in operator (\ref{badguy}).}
\end{figure}
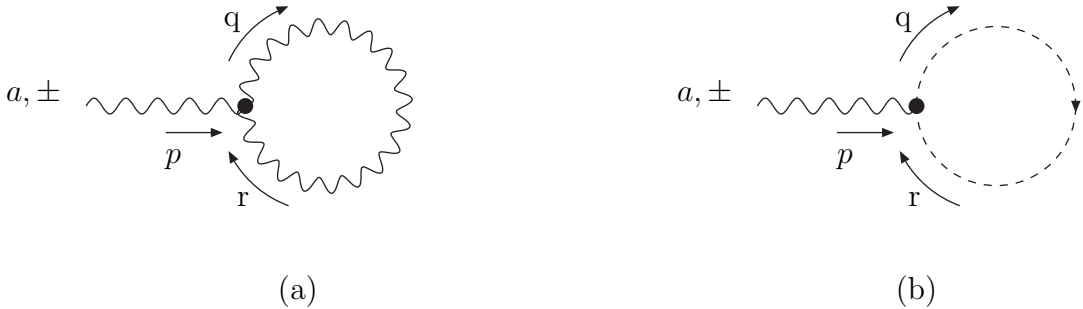

Using the above propagators we can now easily calculate the
contribution to the tadpole (\ref{badguy}). As discussed before, there
can only be a contribution to the $U(1)$ factor $F^8_{+-}$. The
Feynman rule for the gauge three-point function and the
ghost-ghost-gauge coupling are the conventional ones
\begin{eqnarray*}
\begin{picture}(220,110)%(200,110)(-20,0)
%%%%%%%%%%%%%%%%%%%%%%%%%%%%%%%%%%%%%%%%%%%%%%%%%%%%%%%
  \Photon(0,50)(60,50){3}{5} \Vertex(60,50){3}
  \Text(-5,60)[l]{$a, M$}
  \LongArrow(20,40)(40,40)
  \Text(30,30)[c]{$p$}
  \Photon(60,50)(100,100){3}{5}
  \Text(105,100)[l]{$b, N$}
  \LongArrow(90,75)(80,62.5)
  \Text(95,75)[l]{q}
  \Photon(100,0)(60,50){3}{5}
  \Text(105,0)[l]{$c, R$}
  \LongArrow(90,25)(80,37.5)
  \Text(95,25)[l]{r}
%%%%%%%%%%%%%%%%%%%%%%%%%%%%%%%%%%%%%%%%%%%%%%%%%%%%%%%%
\end{picture}
&
\begin{picture}(100,110)%(200,110)(-20,0)
%%%%%%%%%%%%%%%%%%%%%%%%%%%%%%%%%%%%%%%%%%%%%%%%%%%%%%%
  \Photon(0,50)(60,50){3}{5} \Vertex(60,50){3}
  \Text(-5,60)[l]{$a, M$}
  \LongArrow(20,40)(40,40)
  \Text(30,30)[c]{$p$}
  \DashArrowLine(60,50)(100,100){3}
  \Text(105,100)[l]{$b$}
  \LongArrow(90,75)(80,62.5)
  \Text(95,75)[l]{q}
  \DashArrowLine(100,0)(60,50){3}
  \Text(105,0)[l]{$c$}
  \LongArrow(90,25)(80,37.5)
  \Text(95,25)[l]{r}
%  \Text(140,50)[l]{$-g f^{abc}\, q_{M}$}
%%%%%%%%%%%%%%%%%%%%%%%%%%%%%%%%%%%%%%%%%%%%%%%%%%%%%%%%
\end{picture} \\
\\
g f^{abc}  \left[g_{MN} (q-p)_R + g_{NR}(r-q)_M +g_{RM}(p-r)_N\right]
&
g f^{abc}\, q_{M}
\end{eqnarray*}
\hspace{.2cm}
where the structure constants $f^{abc}$ are given by
\begin{equation}
  f^{abc} = -2 i \,{\rm tr} (T^a [T^b,T^c])
  \ \ {\it i.e.} \ \
  [T^b,T^c] = i \, f^{bcd} {\tilde g}_{da} T^a.
\end{equation}
Note that the vertices are conserving the 6D momenta. The violations
of translational invariance in the extra dimensions appears only
through the propagators. By momentum conservation, the in-going 4D
momentum in the tadpole diagrams in Fig.~\ref{fig:tadpole} is
vanishing. The momentum along the extra dimensions circulating in the
loop is related to the in-going one by the delta functions of the
propagator.  Explicitly the gauge tadpole diagram in
Fig.~\ref{fig:tadpole}.a is given by
\begin{equation}
  {\rm TadGa}^a_M
  =
  \sfrac{1}{2} \int \frac{d^4q}{(2\pi)^4}\, g f^{abc}
  ((r+q)_M g^{NR} - (p+q)^R \delta^{N}_{M} + (p-r)^N \delta^{R}_{M})\,
  \langle A^b_N (q) A^c_R (r) \rangle
\end{equation}
Note that the factor $1/2$ is the expression of the tadpole is just a
symmetry factor.  Explicitly evaluating these terms we find that the
only non-vanishing components of the gauge tadpole are (in the $\xi=1$
gauge for the $G_2$ model on $T^2/Z_4$):
\begin{eqnarray}
  {\rm TadGa}^8_+ & = & - g \sqrt{\frac{2}{3}} \int \frac{d^4q}{(2\pi)^4}\
  \frac{p_+}{q_\mu q^\mu -  p_+p_-}
  \nonumber \\
  {\rm TadGa}^8_- & = & g \sqrt{\frac{2}{3}}  \int \frac{d^4q}{(2\pi)^4}\
  \frac{p_-}{q_\mu q^\mu -  p_+p_-}
\end{eqnarray}

The ghost tadpole diagram in Fig.~\ref{fig:tadpole}.b is given by
\begin{equation}
  {\rm TadGh}^a_M
  =
  - \int \frac{d^4q}{(2\pi)^4}\, g f^{abc} (-q_M) \,
  \langle C^b (q) {\bar C}^c (r) \rangle
\end{equation}
Note that the minus sign is due to the anti-commuting nature of the
ghosts.  The only non-vanishing components for the $G_2$ case are:
\begin{eqnarray}
  {\rm TadGh}^8_+ & = &  g \sqrt{\frac{2}{3}} \int  \frac{d^4q}{(2\pi)^4}\
  \frac{p_+}{q_\mu q^\mu -  p_+p_-}
  \nonumber \\
  {\rm TadGh}^8_- & = & - g \sqrt{\frac{2}{3}} \int \frac{d^4q}{(2\pi)^4}\
  \frac{p_-}{q_\mu q^\mu -  p_+p_-}
\end{eqnarray}
The gauge and ghost loops exactly cancel each other and no tadpole is
generated at one-loop.

Note that in the $SU(3)$ model, the same result holds with slightly
different numerical coefficients, the factor $\sqrt{2/3}$ being
replaced by $\sqrt{3}/2$.

 Due to the power law running of the gauge couplings, the cutoff
scale  of the 6D theory cannot be pushed much higher than a few times 10~TeV.
Therefore, even if generated at higher loop, the tadpole operators will be
phenomenologically harmless. From a theoretical point of view, however, it 
would
be very interesting to know if such operators are generated at any
perturbative level.

%%%%%%%%%%%%%%%%%%%%%%%%%%%%%%%%%%%%%%%%%%%%%%%%%%%%%
%%%%%%%%%%%%%%%%%%%%%%%%%%%%%%%%%%%%%%%%%%%%%%%%%%%%%
\section{Introducing Fermions and Yukawa Couplings}
\setcounter{equation}{0}
\setcounter{footnote}{0}
%%%%%%%%%%%%%%%%%%%%%%%%%%%%%%%%%%%%%%%%%%%%%%%%%%%%%
%%%%%%%%%%%%%%%%%%%%%%%%%%%%%%%%%%%%%%%%%%%%%%%%%%%%%

The fermion sector has been the common difficulty with the Higgs boson
originating in higher-dimensional gauge bosons.  In the original
incarnation by Manton~\cite{Manton}, the chiral fermions could not be
obtained because the group $G_2$ is real.  In general, obtaining the
correct standard model fermion content is a challenge.  Another
problem is that the Higgs doublet is a higher-dimensional gauge boson,
and hence its coupling is dictated by the gauge symmetry.  It appears
arbitrary Yukawa couplings are not allowed.

The main new ingredient in our model is the orbifold.  At the orbifold
fixed point, we can introduce fermions that transform only under the
unbroken $SU(2) \times U(1)$ rather than the full $G_2$.  Therefore we
can introduce the correct fermion content of the standard model
without difficulty.  Once the fermions are at the fixed point, their
Yukawa couplings are not directly tied to the gauge interactions in
the bulk.  Using the Wilson line operator, we can now write arbitrary
Yukawa couplings we need.

%%%%%%
\subsection{Yukawa couplings from Wilson line interactions}

We have seen that one can build a successful model of the bosonic
sector of the SM based on extra dimensional gauge theories.
This sector has no one loop quadratic divergences, and the zero
modes reproduce the bosonic matter content of the SM plus a single
scalar doublet with hypercharge 3/2. In principle there are two
possibilities for introducing fermions: they can be in the bulk or at
the orbifold fixed points. Since the Higgs is part of the extra
dimensional gauge field, then if fermions are introduced in the bulk
their Yukawa couplings will be determined by the bulk $G_2$ gauge
coupling, and the Yukawa couplings for the different families will be
equal. Thus it seems very difficult to obtain a realistic fermion mass
pattern this way. Therefore the fermions should be introduced at the
fixed points. This is a generic conclusion for models where the Higgs
is part of the extra dimensional gauge field. In the particular case
at hand, there is another reason why the SM fermions should be at the
fixed point: since the embedding of the SM into $G_2$ is via the
$SU(3)$ subgroup of $G_2$ we hit the usual problem of embedding quarks
into $SU(3)$: their hypercharges are fractional with respect to the
hypercharge unit of $SU(3)$, so there is no representation that would
give the correct $U(1)_Y$ quantum numbers.

Once the SM fermions are introduced at the orbifold fixed points, one
could try to directly linearly couple the SM fermions to the Higgs
field at the fixed point. However this clearly reintroduces the
quadratic divergences already at the one loop level, and is clearly
not a desirable solution. Also introducing Yukawa couplings this way
explicitly breaks the shift symmetry, $A_i^{\hat{a}}\to A_i^{\hat{a}}
+\partial_i \xi^{\hat{a}}$  in its infinitesimal form
(${\hat a}$ corresponds a broken generator
index and $i=y,z$), which is the remnant of higher dimensional gauge
invariance at the fixed points. Thus one would like to look for
operators that do not break this shift symmetry. This can be achieved
by using operators that involve Wilson lines between the fixed points
(the two fixed points may also coincide) $W=P e^{i\int A_i dx^i}$.
Since the gauge transformation parameter for the broken generators
vanishes at the fixed points $\xi^{\hat{a}}|_{fp}=0$, the Wilson line
will be invariant under this symmetry. Inspired by this observation we
will construct interaction terms containing Wilson lines. We will
require the cancellation of one-loop quadratic divergences from the
newly induced couplings.

One may wonder where such Yukawa couplings involving Wilson lines
could originate from.  It is worth recalling that fermions at the
fixed points can arise in the twisted sector in string theory.  They
are ``localized'' because the string winds around the fixed point, and
are therefore not strictly at the fixed {\it point}.  They are spread
out around the fixed point for a finite distance.  If this spreading
is large enough, the ``wave functions'' for states at different fixed
points can overlap and can have couplings.  For instance,
Ref.\cite{Ibanez} tried to generate Yukawa hierarchy using states at
different fixed points.  In the low-energy effective field theory
description of couplings of twisted-sector states at different fixed
points, the gauge invariance requires that the couplings come together
with the Wilson lines.  Therefore we expect that non-local
interactions with Wilson lines are natural in this context.

It is an interesting question if we can generate Yukawa couplings with
Wilson lines in a purely field-theoretical construction.  We will
argue that it is natural to expect the appearance of these operators
once some massive bulk fermions which could mix with fields at the
fixed points are integrated out. 
To illustrate this, consider a simple example with a
single extra dimension compactified on a circle.  Assume that there is
a massive 5D fermion $\psi$ living in the bulk, and that a constant
$A_y$ gauge field is turned on. The fermion propagator in the presence
of the constant gauge field is just
\begin{equation}
  \langle \bar\psi (p) \psi (p) \rangle  
  = \frac{i}{p_\mu \gamma^\mu -p_y \gamma^y+gA_y \gamma^y-m -i\epsilon} .
\end{equation}
The quantization condition for $p_y$ will be $p_y=\frac{2\pi n}{L}+g
A_y$.  The propagator in coordinate space along the extra dimension
will then be
\begin{equation}
  \langle \bar\psi (p_\mu,y_1) \psi (p_\mu,y_2) \rangle  
  = \sum_{p_y=\frac{2\pi n}{L}+g A_y} e^{i p_y(y_1-y_2)}
  \frac{i}{p_\mu \gamma^\mu -p_y \gamma^y+gA_y \gamma^y -m -i\epsilon} .
\end{equation}
Shifting the summation to $p_y'=p_y-gA_y$ we get that
\begin{equation}
  \langle \bar\psi (p_\mu,y_1) \psi (p_\mu,y_2) \rangle _{A_y}=
  e^{ig A_y (y_2-y_1)} \langle \bar\psi (p_\mu,y_1) \psi (p_\mu,y_2)
  \rangle _{A_y=0}.
\end{equation}
Thus the Wilson line appears in the propagator. If there are
couplings of the form
\begin{equation}
  \int d^5x \left( \bar{\psi} (D_A \gamma^A -m) \psi
    + \delta (y_1) \bar{\psi} (y_1) \chi (y_1) 
    +\delta (y_2) \bar{\xi} (y_2) \psi (y_2)
\right),
\end{equation}
a non-local interaction term of the form
\begin{equation}
  \label{eq:ExampleW}
  W=\int d^4 x \, C\, \bar{\xi}(y_2) e^{ig \int_{y_1}^{y_2} A_y dy} \chi (y_1)
\end{equation}
will be generated. Thus one would expect that such operators could
generically appear in a theory after the massive fermions are
integrated out. However, since it involves the massive fermion
propagator, the coefficient $C$ of this will include the suppression
factor $e^{-m |y_1-y_2|}$. Therefore in order to get a sizable
coupling the bulk fermion should not be much heavier than the inverse
radius of the extra dimension. On the other hand, the exponential
factor could be used to generate fermion mass hierarchies by varying
the bulk mass of the fermion that is being integrated out. This
procedure of integrating out heavy fermions thus can give the
operators needed to generate the Yukawa couplings for the leptons.
Note that this is nothing but the Froggatt--Nielsen
mechanism~\cite{FN} except that the summation over the entire KK tower
of the ``Froggatt--Nielsen fermion'' gives an exponential rather than
a power suppression.  Because the generated Yukawa couplings depend
exponentially on the mass of the bulk fermion, it is easy to generate
a large hierarchy among Yukawa couplings.  We find this an attractive
mechanism to explain the fermion mass hierarchy.  Moreover, the mass
of the bulk fermion is protected by chiral symmetry, and hence the
radiative correction to the fermion mass is proportional to the bare
mass.  Therefore this mechanism is technically natural.

However, for the quarks there is an added difficulty due to the
fractional $U(1)_Y$ charges of the quarks. Such fields can not mix
with bulk fermions, and therefore some other mechanism is needed to
generate the Wilson line interactions.

A comment on the gauge anomaly is in order.  When left-handed and
right-handed fermions are split on different fixed points, the
four-dimensional gauge anomaly is not canceled at each fixed point.
It requires the anomaly flow from one fixed point to the other.  This
can be easily done by integrating the five-dimensional Chern--Simon
term from one fixed point to the other.  It is well-known that the
gauge variation of the Chern--Simon term is a total derivative, whose
surface term precisely gives the four-dimensional Wess--Zumino
consistent anomaly (see, e.g., Ref.\cite{Zumino}).  This is not an
accident; it is a direct consequence of family's index
theorem~\cite{AGG}. Note that bulk massive fermions are vector-like by
definition and do not contribute to six-dimensional nor
four-dimensional gauge anomalies.

%%%%%%
\subsection{One-loop radiative corrections to Higgs mass from Yukawa couplings}

The Wilson line transforms as a fundamental under the gauge group
$G_2$ at the starting point of the integration, and as an
anti-fundamental under the gauge group at the endpoint of integration.
If the starting and ending points coincide then the Wilson line will
be in the adjoint representation. Of course since in our case the
endpoints of integration are the orbifold fixed points, and only the
$SU(2)\times U(1)$ subgroup of $G_2$ is active at the fixed points,
the Wilson line also only transforms under $SU(2)\times U(1)$ at the
fixed points. The Wilson line can be represented as a $7\times 7$
matrix $W^a_b = P e^{i\int A_i dx^i}$. We arrange one generation of
quarks into seven component vectors of the form
\begin{equation}
  Q_L= \left( \begin{array}{c} u_L \\ d_L \\ 0 \\ 0 \\ 0\\ 0\\ 0\end{array}
  \right) = P_1 \tilde{q}_L, \ \ \ U_R= \left( \begin{array}{c}
      0 \\ 0  \\ 0 \\ 0 \\ 0\\ u_R\\ 0\end{array}
  \right) =P_2 \tilde{u}_R, \ \ \  D_R= \left( \begin{array}{c}
      0 \\ 0  \\ 0 \\ 0 \\ 0\\ 0\\ d_R \end{array}
  \right) =P_3 \tilde{d}_R,
\end{equation}
where the $P_i$ projection operators are
\begin{equation}
  \label{projectors}
  P_1={\rm diag} (1,1,0,0,0,0,0), \ \ P_2={\rm diag} (0,0,0,0,0,1,0), \ \
  P_3={\rm diag} (0,0,0,0,0,0,1).
\end{equation}
At this point it should be noted that the quarks transform in the
usual way under the unbroken $SU(2)$ symmetry at the fixed point. As
far as the hypercharges are concerned however, the naive action of
$\sqrt{6}T^8$ would not give the SM numbers (we are saying nothing but
the fact that the quarks, because of their fractional charges, cannot
be embedded into full representations of $G_2$). Fortunately, we are
free to define the quarks hypercharges as we want and therefore we
will assign them their SM values, which will allow us to construct
$SU(2)\times U(1)$ invariant interactions.

Let us consider the interaction term of the form
\begin{equation}
                                %        \label{eq:LYuk}
  {\cal L}_{\rm \scriptstyle Yukawa}
  = m_u {\bar U}_R  W_u Q_L +
  m_d {\bar D}_R  W_d Q_L.
\end{equation}
The one loop quadratic divergence from the Coleman-Weinberg formula is
then given by
\begin{eqnarray}
  &&
  \frac{3\Lambda^2}{16\pi^2}
  \left (m_u^2 {\rm Tr}( P_2 W_u P_1 P_1 W_u^\dagger P_2 )
    + m_d^2 {\rm Tr} (P_3 W_d P_1 P_1 W_d^\dagger P_3) \right)
  \nonumber\\
  &&
  \label{eq:ColemanW}
  \hspace{2cm}
  = \frac{3\Lambda^2}{16\pi^2}
  \left(
    m_u^2 {\rm Tr} (P_2 W_u P_1W_u^\dagger)
    +m_d^2 {\rm Tr} (P_3 W_d P_1W_d^\dagger)
  \right).
\end{eqnarray}
Thus the quadratic divergences will cancel if either the projector
$P_1$ or $P_{2,3}$ commutes with the Wilson line $W$. Since we only
want to avoid quadratic divergences for the hypercharge 1/2 Higgs, the
requirement really is that the projector should commute with the
matrix in (\ref{Higgs1})--(\ref{Higgs2}) with $H\to 0$.  This is not
true for the projectors in (\ref{projectors}), but can be fixed
analogously to the mechanism employed in little Higgs
theories~\cite{littlehiggs} by introducing more fermions at the fixed
point, that is by filling more of the diagonal components of $P_1$ or
of $P_2$ and $P_3$. The simplest possibility is to fill $P_1$ to be
the identity matrix. In this case it clearly will commute with $W$ and
there will be {\it no quadratic divergences for any fields} but only a
contribution to the vacuum energy.  The origin of the cancellation of
the quadratic divergences will then be the chiral symmetry between the
newly introduced color triplet fermions and the doublet
quarks~\cite{littlehiggs}.

However, from a phenomenological point of view, one can get away from
dangerous divergences by introducing fewer fields.  In fact,
introducing a single color triplet field in the third family $\chi_L$
such that $P_1={\rm diag}(1,1,0,0,0,1,0)$ suffices.  Indeed, by
examination of ~(\ref{eq:ColemanW}), we get that the quadratic
divergences in the SM Higgs mass from the top Yukawa coupling do
cancel; we are left with quadratic divergences from the bottom Yukawa
coupling which, phenomenologically, are harmless due to the smallness
of the coupling.  We also get, from the top sector, a (positive)
quadratically divergent correction to the square mass of the
hypercharge 3/2 Higgs, which is good since it will push its mass close
to the cutoff scale of the theory and will prevent him from getting a
VEV.  In conclusion, we are going to consider
\begin{equation}
  Q_L= \left( \begin{array}{c} u_L \\ d_L \\ 0 \\ 0 \\ 0\\ 
      \chi_L\\ 0\end{array}
  \right), \ \ \ U_R= \left( \begin{array}{c}
      0 \\ 0  \\ 0 \\ 0 \\ 0\\ u_R\\ 0\end{array}
  \right) , \ \ \  D_R= \left( \begin{array}{c}
      0 \\ 0  \\ 0 \\ 0 \\ 0\\ 0\\ d_R \end{array}
  \right) ,
\end{equation}
along with the interaction term:
\begin{equation}
  \label{eq:LYuk}
  {\cal L}_{\rm \scriptstyle Yukawa}
  = m_u {\bar U}_R  W_u Q_L +
  m_d {\bar D}_R  W_d Q_L
  +M {\bar \chi}_R \chi_L.
\end{equation}
%

%%%%%%
\subsection{Explicit Computation of One-loop radiative corrections to
  Higgs mass}

\begin{figure}
\centerline{
\hspace{1cm}
\begin{picture}(150,110)%(110,110)(-20,0)
%%%%%%%%%%%%%%%%%%%%%%%%%%%%%%%%%%%%%%%%%%%%%5
 \DashArrowLine(0,20)(50,20){3}
  \Vertex(50,20){3}
  \Text(5,10)[l]{$h$}
  \DashArrowLine(50,20)(100,20){3}
  \Text(95,10)[l]{$h$}
  \ArrowArcn(50,50)(30,270,90)
  \Text(10,50)[r]{$R_{\rm \scriptstyle heavy}$}
  \ArrowArcn(50,50)(30,90,270)
  \Text(90,50)[l]{$R_{\rm \scriptstyle heavy}$}
%%%%%%%%%%%%%%%%%%%%%%%%%%%%%%%%%%%%%%%%%%%%%%%%%%%%%%%%
\end{picture}
\hspace{4cm}
\begin{picture}(150,110)%(110,110)(-20,0)
%%%%%%%%%%%%%%%%%%%%%%%%%%%%%%%%%%%%%%%%%%%%%5
  \DashArrowLine(-20,50)(20,50){3}
  \Vertex(20,50){3}
  \Text(-15,40)[l]{$h$}
  \DashArrowLine(80,50)(120,50){3}
  \Text(115,40)[l]{$h$}
  \Vertex(80,50){3}
  \ArrowArcn(50,50)(30,180,0)
  \Text(50,93)[c]{$Q_L$}
  \ArrowArcn(50,50)(30,0,180)
  \Text(50,5)[c]{$R_{\rm \scriptstyle heavy}$, $R_{\rm \scriptstyle light}$}
%%%%%%%%%%%%%%%%%%%%%%%%%%%%%%%%%%%%%%%%%%%%%%%%%%%%%%%%
\end{picture}
}
\centerline{
\hspace{.8cm}
\begin{picture}(150,110)%(110,110)(-20,0)
%%%%%%%%%%%%%%%%%%%%%%%%%%%%%%%%%%%%%%%%%%%%%5
 \DashArrowLine(0,20)(50,20){3}
  \Vertex(50,20){3}
  \Text(5,10)[l]{$H$}
  \DashArrowLine(50,20)(100,20){3}
  \Text(95,10)[l]{$H$}
  \ArrowArcn(50,50)(30,270,90)
  \Text(10,50)[r]{$R_{\rm \scriptstyle heavy}$}
  \ArrowArcn(50,50)(30,90,270)
  \Text(90,50)[l]{$R_{\rm \scriptstyle heavy}$}
%%%%%%%%%%%%%%%%%%%%%%%%%%%%%%%%%%%%%%%%%%%%%%%%%%%%%%%%
\end{picture}
}
\caption{Radiative corrections, from Yukawa interactions,
  to the SM Higgs square mass, $m_h^2$, and to the $3/2$ hypercharge
  Higgs square mass, $m_H^2$.}
\label{fig:Yuk}
\end{figure}
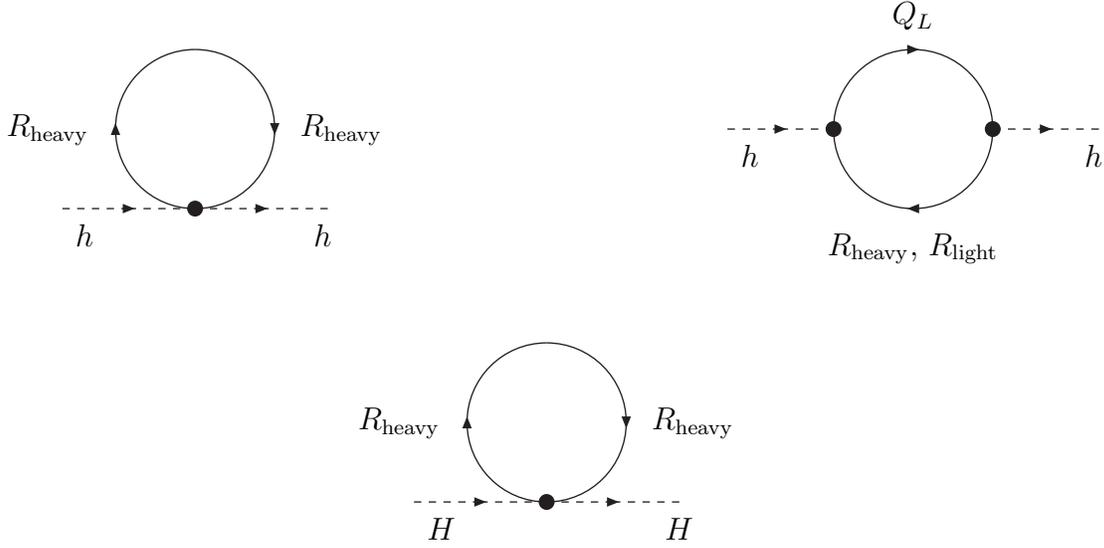

We want to explicitly compute the radiative corrections to the Higgs
mass from Yukawa interactions.  We need to expand the action
(\ref{eq:LYuk}) up to quartic order and for simplicity we will retain
only the top Yukawa coupling (we assume without loss of generality
that $m_u$ is real):
\begin{eqnarray}
  &&
  {\cal L}_{\rm \scriptstyle Yukawa}  =
  m_u {\bar u}_R \chi_L + M {\bar \chi}_R \chi_L
  + {\tilde \lambda}_u  {\bar u}_R q_L h
  -\frac{|{\tilde \lambda}_u|^2}{2m_u} {\bar u}_R \chi_L (3 |H|^2 +|h|^2)
\end{eqnarray}
where the Yukawa coupling ${\tilde \lambda}_u$ is obtained from the expansion of the Wilson line
interaction~(\ref{eq:ExampleW}), {\it i.e.} ${\tilde \lambda}_u \sim C m_u g$.
The right handed up quark mixes with the extra fermion $\chi$ and
becomes massive.  The mass eigenstates are:
\begin{eqnarray}
  & \displaystyle
  R_{\rm \scriptstyle light} 
  = \frac{1}{\sqrt{m_u^2+M^2}} ( M u_R - m_u \chi_R )
  \ \ \ {\rm with} \ \ \ m=0;\\
  & \displaystyle
  R_{\rm \scriptstyle heavy} 
  = \frac{1}{\sqrt{m_u^2+M^2}} ( m_u u_R + M \chi_R )
  \ \ \ {\rm with} \ \ \ m^2=m_u^2+M^2.
\end{eqnarray}
Then the action becomes:
\begin{eqnarray}
  {\cal L}_{\rm \scriptstyle Yukawa}  =
  \sqrt{\lambda_u^2+M^2} {\bar R}_{\rm \scriptstyle heavy} \chi_L
  + \frac{{\tilde \lambda}_u m_u}{\sqrt{m_u^2 + M^2}} 
  {\bar R}_{\rm \scriptstyle heavy}q_L h
  + \frac{{\tilde \lambda}_u M}{\sqrt{m_u^2 + M^2}} 
  {\bar R}_{\rm \scriptstyle light}q_L h
  \nonumber\\
  \label{eq:EffYuk}
  -\frac{|{\tilde \lambda}_u|^2}{2\sqrt{m_u^2+M^2}} 
  {\bar R}_{\rm \scriptstyle heavy} \chi_L (3 |H|^2 + |h|^2)
  -\frac{|{\tilde \lambda}_u|^2 M}{2m_u\sqrt{m_u^2+M^2}} 
  {\bar R}_{\rm \scriptstyle light} \chi_L (3 |H|^2 + |h|^2)
\end{eqnarray}
The diagrams contributing to the Higgs masses at one loop are depicted
on Fig.~\ref{fig:Yuk}.
Computing the diagrams and including the color factor for the
fermions, we get :
\begin{eqnarray}
  \label{eq:mh2}
  & \displaystyle
  \delta m_h^2   =
  6 i |{\tilde \lambda}_u|^2
  \int \frac{d^4q}{(2\pi)^4}
  \left( \frac{M^2}{q^2 (q^2 - m_u^2 - M^2)} \right)
  = -\frac{3|{\tilde \lambda}_u|^2 M^2}{8 \pi^2}
  \ln \left( \frac{\Lambda^2 + m_u^2 + M^2}{m_u^2 + M^2} \right);
  \\
  \label{eq:mH2}
  & \displaystyle
  \delta m_H^2  =   i \sfrac{9}{2}  |{\tilde \lambda}_u|^2
  \int \frac{d^4q}{(2\pi)^4} \frac{1}{q^2-m_u^2-M^2}
  = \frac{3 |{\tilde \lambda}_u|^2}{32 \pi^2} \Lambda^2
  +\ldots .
\end{eqnarray}
As in softly broken supersymmetric theories, the one-loop radiative
corrections to the Higgs mass square from the top Yukawa coupling are
negative and trigger the Electroweak symmetry breaking.  Furthermore,
we see that the radiative corrections to the hypercharge 3/2 Higgs
mass square are quadratically divergent and positive. This ensures
that this scalar doublet will not acquire a VEV and decouples from the
low-energy effective theory.

%%%%%%%%%%%%%%%%%%%%%%%%%%%%%%%%%%%%%%%%%%%%%%%%%%%%%
%%%%%%%%%%%%%%%%%%%%%%%%%%%%%%%%%%%%%%%%%%%%%%%%%%%%%
%%%%%%%%%%%%%%%%%%%%%%%%%%%%%%%%%%%%%%%%%%%%%%%%%%%%%
\subsection{Estimates for the Scales of the Theory}

The parameters of the theory relevant to electroweak symmetry breaking
are $R$, the radius of the orbifold, $\lambda_u$, the coefficient of
the Yukawa coupling, $M$ and $m_u$, the mass parameters for the
colored fermions, and the cutoff scale $\Lambda$. In order to estimate
the size of these parameters we need to calculate the effective Higgs
potential. It will have several contributions
\begin{eqnarray}
  V_{\it eff} (h)&=&  -\frac{3|{\tilde \lambda}_u|^2 M^2}{8 \pi^2}
  \ln \left( \frac{\Lambda^2 + m_u^2 + M^2}{m_u^2 + M^2} \right) h^\dagger h
  +m_{bulk}^2 h^\dagger h + C \frac{|{\tilde \lambda}_u|^4\Lambda^2}{(16
    \pi^2)^2} h^\dagger h
  \nonumber \\ &&
  +\frac{1}{6} g^2 (h^\dagger h)^2.
  \label{Veff}
\end{eqnarray}
Here the first term is the one-loop contribution from the Yukawa
sector calculated in the previous section. $m_{bulk}^2$ is the finite
contribution to the scalar masses, which have to be calculated for
this particular orbifold.  In 5D orbifold theories such contributions
have been calculated in \cite{ABQ,GIQ,Martin}, and are of the order $9
g^2 \zeta (3) C_2 (G)/(32 \pi^2 R^2)$. The final mass term is the
two-loop quadratic divergence that is expected to appear due to the
Yukawa sector. The quartic scalar potential appears from the bulk
gauge interactions, and gets a logarithmic running from the Yukawa
couplings and from brane induced pieces from gauge interactions. The
bulk contribution reduces the size of the negative Higgs mass term
from the Yukawa couplings, while the sign of the two-loop contribution
would have to be explicitly calculated. Since
\begin{equation}
  1 \sim \lambda_{\rm \scriptstyle top}
  =  \frac{M {\tilde \lambda}_u}{\sqrt{m_u^2+M^2}};
\end{equation}
and we expect $M \sim m_u$, therefore ${\tilde \lambda}_u \sim 1$.  In
order to get the correct electroweak symmetry breaking VEV for the
Higgs, we would need the minimum of the Higgs potential to be at
$\langle h^\dagger h \rangle = v^2/2$. Thus
\begin{equation}
  \frac{3}{8\pi^2} M^2
  \ln \left( \frac{\Lambda^2 + m_u^2 + M^2}{m_u^2 + M^2} \right)
  -m_{bulk}^2  -  C \frac{|{\tilde \lambda}_u|^4\Lambda^2}{(16
    \pi^2)^2} \sim \frac{2}{3} M_W^2, 
\end{equation}
and so $M\sim 5 - 10 \times M_W$.  Therefore one would expect the
relevant scales of the theory to be in the range $1$~TeV $\gsim
M,m_u,1/R \gsim 500$~GeV.  The cutoff scale of the theory could then
be a factor of 10-20 larger than the mass scale of these particles and
thus of the order $5-20$~TeV. Note that the necessary scale for new
physics is quite low, and therefore a detailed analysis should be
performed to determine which region of the parameter space could be
consistent with all experimental constraints.

The particle spectrum of this theory would then be as follows. Below
the characteristic scale $1/R \sim M \sim 500$~GeV--$1$~TeV, one would
only have the SM particles. The Higgs mass should be estimated from
(\ref{Veff}). Since the quartic scalar coupling is fixed by the gauge
couplings (similar to supersymmetric models), the Higgs is expected to
be light. By minimizing (\ref{Veff}) the value of the Higgs mass using
the tree-level quartic scalar coupling would be
$m_h^{\rm \scriptstyle tree}=\sqrt{\frac{4}{3}} M_W = M_Z \sim 91$~GeV (to the extent
that we use the approximate prediction $\sin^2 \theta_W = 1/4$).
The loop corrections to the quartic scalar couplings from the Yukawa
sector and also from the gauge sector will result in additional
contributions.  For example, from the top Yukawa coupling one gets a
correction to the quartic scalar coupling of order
\begin{equation}
  \delta V (h) \sim
  - \frac{ 3 \lambda_{\rm \scriptstyle top}^4}{16 \pi^2} \ln (m_h  R)^2 \ (h^\dagger h)^2
  \ > 0,
\end{equation}
which itself would raise the Higgs mass to $\sim 130$~GeV.  One
generically expects the Higgs to be much below the $500$~GeV--$1$~TeV
scale, in the 120--150~GeV regime, and likely within the reach of
Tevatron Run II. Note that the zero mode of the second Higgs doublet
with hypercharge 3/2 does get quadratically divergent corrections due
to the structure of the Yukawa sector, and thus its mass is expected
to be of order few $\Lambda^2/16 \pi^2 \sim $TeV--few~TeV.  Once we
get to the scale $1/R \sim 500$~GeV--$1$~TeV we will start exploring
the KK spectrum of the bosonic modes. In particular, the KK modes of
the full $G_2$ gauge boson sector should appear. From the $A_y,A_z$
sector it is likely that just like the hypercharge 3/2 Higgs most
states will get quadratically divergent mass contributions from the
Yukawa sector and their KK towers thus will start at a scale higher
than those for the gauge fields, except for the physical SM Higgs
itself, which as we saw above is much lighter than $1/R$.  Of course
some of these states will just serve as longitudinal modes for the
massive KK gauge bosons.  Also around the $1/R\sim M\sim
500$~GeV--$1$~TeV scale the colored fermions $\chi,\bar{\chi}$ needed
to cancel the divergences from the Yukawa couplings for the Higgs will
show up. Thus the bosonic sector of this theory is that of an extra
dimensional model, while the fermion sector would much look like that
of a little Higgs model~\cite{littlehiggs,minimallittle,otherlittle}.
This is due to the construction of the model, where all bosons come
from bulk gauge fields, while since the fermions are introduced at the
orbifold fixed point, their description is essentially equivalent to
that of the little Higgs models.

%%%%%%%%%%%%%%%%%%%%%%%%%%%%%%%%%%%%%%%%%%%%%%%%%%%%%
%%%%%%%%%%%%%%%%%%%%%%%%%%%%%%%%%%%%%%%%%%%%%%%%%%%%%
\section{Conclusions}
\setcounter{equation}{0}
\setcounter{footnote}{0}
%%%%%%%%%%%%%%%%%%%%%%%%%%%%%%%%%%%%%%%%%%%%%%%%%%%%%
%%%%%%%%%%%%%%%%%%%%%%%%%%%%%%%%%%%%%%%%%%%%%%%%%%%%%

We have considered the possibility that the standard model Higgs
originates from a 4D scalar component of a higher dimensional gauge
field. In this case higher dimensional gauge invariance could protect
the Higgs from some of the quadratically divergent loop corrections
that plague the Standard Model. We have considered orbifold
compactifications of higher dimensional gauge theories, and found that
the preferred model is a 6D $G_2$ gauge theory compactified on a $Z_4$
(or $Z_{k\geq 4}$) orbifold, where the orbifold breaks the bulk $G_2$
gauge group down to $SU(2)\times U(1)$. This model would predict a
value of $\sin^2\theta_W =1/4$, after the zero mode of one of the
scalar components of the 6D gauge field is identified with the SM
Higgs.

One needs to check whether in such models the orbifold projection
itself would reintroduce the quadratic divergences on the fixed
points.  We have found that in general for $Z_2$ compactifications
such divergences (and the tadpole operators they would accompany) are
forbidden by the parity invariance of the gauge sector, however for
higher $Z_k$  we needed to explicitly compute one loop diagrams to see that it
is vanishing. Thus the bosonic sector
of this model can accommodate the SM without any one-loop
quadratic divergences.

It had been more difficult to incorporate fermion fields. Since one wants to
have the option of generating different Yukawa couplings for the
different generations, the SM fermions need to be introduced at the
fixed points. Another reason for this is that quarks have fractional
hypercharge quantum numbers in the unit dictated by the bulk gauge
group.  In order to maintain the symmetries of the bulk one then needs
to add Yukawa couplings in the form of non-local Wilson lines, which
generically can be obtained by integrating out bulk fermions that mix
with the brane fields. In order to cancel the one-loop quadratic
divergence for the Higgs from the Yukawa sector additional massive
fermions need to be added to the orbifold~fixed~point.

These theories generically predict a light Higgs boson, since the
quartic scalar coupling is related to the gauge coupling, just like in
the MSSM. The bosonic sector of these models would give KK towers to
all bulk gauge fields, starting at $1/R \sim 500$~GeV--1~TeV, while
the fermionic sector would resemble those of the little Higgs models.

As for the full resolution to the hierarchy problem, there are several
obvious issues to be resolved. The Yukawa sector yields higher loop
quadratic divergences. There could also be non-perturbative
corrections in the strongly interacting higher dimensional theory of
order $\Lambda^2 e^{-\Lambda R}$, which could be as large as the
two-loop quadratic divergences themselves.  Finally, one would have to
explain, why the radion field, which will appear once gravity in 6D is
made dynamical, would be stabilized at the right minimum of order
$\sim {\rm TeV}^{-1}$.

%%%%%%%%%%%%%%%%%%%%%%%%%%
\section*{Acknowledgments}
We thank Nima Arkani-Hamed, Lawerence J. Hall, Yasunori Nomura, Luigi
Pilo, and Mariano Quir\'os for useful discussions.  We also thank
Mariano Quir\'os for sharing a draft of \cite{GIQ2} with us prior to
publication. We are grateful to Hsin-Chia Cheng for pointing out a mistake in
the all-order proof for the absence of the
tapdople in the first version of this paper.
We are grateful to the Aspen Center for Physics, where
this work was initiated during the workshop ``Advances in Field Theory
and Applications to Particle Physics." C.C. and C.G. thank the T-8
group of Los Alamos National Laboratory for providing a stimulating
environment during the Santa~Fe Institute 2002.  C.C. is supported in
part by the DOE OJI grant DE-FG02-01ER41206 and by the NSF grant
PHY-0139738.  C.G. is indebted to the Argonne HEP Theory Group for its
hospitality and its financial support from the United States
Department of Energy, Division of High Energy Physics under contract
W-31-109-ENG-38.  C.G. was supported in part by the RTN European
Program HPRN--CT--2000--00148 and the ACI Jeunes Chercheurs 2068.
H.M. was supported in part by the United States Department of Energy,
Division of High Energy Physics under contract DE-AC03-76SF00098 and
in part by the National Science Foundation grant PHY-0098840.

%%%%%%%%%%%%%%%%%%%%%%%%%%%%%%%%%%%%%%%%%%%%%%%

\appendix

\section*{Appendix}

%%%%%%%%%%%%%%%%%%%%%%

%%%%%%%%%%%%%%%%%%%%%%%%%%%%%%%%%%%%%%%%%%%%%%%%%%%%%
%%%%%%%%%%%%%%%%%%%%%%%%%%%%%%%%%%%%%%%%%%%%%%%%%%%%%
\section{Matrix Representation of $G_2$}
\label{App:algebra}
\setcounter{equation}{0}
\setcounter{footnote}{0}
%%%%%%%%%%%%%%%%%%%%%%%%%%%%%%%%%%%%%%%%%%%%%%%%%%%%%
%%%%%%%%%%%%%%%%%%%%%%%%%%%%%%%%%%%%%%%%%%%%%%%%%%%%%

In this appendix, we give a matrix representation of the fundamental
representation of $G_2$ exhibiting explicitly the $SU(3)$ embedding.
The fundamental being of dimension 7 and the adjoint of dimension 14,
we need fourteen $7\times 7$ matrices:
\begin{eqnarray}
  \label{eq:SU3inG2}
  &
  T^a  = \frac{1}{2\sqrt{2}} \left(
    \begin{array}{ccc}
      \lambda^a  \\
      & -{\lambda^a}^t \\
      & &
    \end{array} \right)
  \ \ \ {\rm for} \ \ \ a=1\ldots 8
  \\
  &
  \hspace{-.2cm}
  T^9
  =
  \frac{i}{2\sqrt{3}} \left(
    \begin{array}{ccc}
      & &  v^1 \\
      \lambda^7   \\
      & {v^1}^\dagger
    \end{array} \right) \
  T^{10}
  =
  \frac{-i}{2\sqrt{3}} \left(
    \begin{array}{ccc}
      & &  v^2 \\
      \lambda^5   \\
      & {v^2}^\dagger
    \end{array} \right) \
  T^{11}
  =\frac{i}{2\sqrt{3}} \left(
    \begin{array}{ccc}
      & &  v^3 \\
      \lambda^2   \\
      & {v^3}^\dagger
    \end{array} \right) \
  \\
  \nonumber \\
  &
  T^{12} = {T^{9}}^\dagger \  \
  T^{13} = {T^{10}}^\dagger \  \
  T^{14} = {T^{11}}^\dagger ,
\end{eqnarray}
where the $\lambda^a$ are the usual Gell-Mann matrices:
\begin{eqnarray}
  \scriptstyle
  &
  \lambda^1
  =
  \left(
    \begin{array}{ccc}
      \sz \oo & \sz 1 & \sz \oo  \\
      \sz 1 &\sz \oo & \sz \oo  \\
      \sz \oo &\sz  \oo & \sz \oo
    \end{array} \right)
  \lambda^2
  =
  \left(
    \begin{array}{ccc}
      \sz \oo & \sz -i & \sz \oo  \\
      \sz i & \sz \oo &\sz  \oo  \\
      \sz  \oo &  \sz \oo &  \sz \oo
    \end{array} \right)
  \lambda^3
  =
  \left(
    \begin{array}{ccc}
      \sz 1 &  \sz \oo &  \sz \oo  \\
      \sz \oo &  \sz  -1 &  \sz \oo  \\
      \sz  \oo &  \sz \oo &  \sz \oo
    \end{array} \right)
  \lambda^8
  =
  \frac{1}{\sqrt{3}}
  \left(
    \begin{array}{ccc}
      \sz 1 &  \sz \oo &  \sz \oo  \\
      \sz \oo &  \sz 1 & \sz  \oo  \\
      \sz \oo &  \sz \oo &  \sz -2
    \end{array} \right)
  \\
  &
  \lambda^4
  =
  \left(
    \begin{array}{ccc}
      \sz \oo & \sz  \oo &  \sz 1  \\
      \sz 1 &  \sz \oo &  \sz \oo  \\
      \sz \oo &  \sz \oo &  \sz \oo
    \end{array} \right)
  \lambda^5
  =
  \left(
    \begin{array}{ccc}
      \sz \oo &  \sz \oo &  \sz -i  \\
      \sz \oo &  \sz \oo &  \sz \oo  \\
      \sz i &  \sz \oo &  \sz \oo
    \end{array} \right)
  \lambda^6
  =
  \left(
    \begin{array}{ccc}
      \sz \oo &  \sz \oo &  \sz \oo  \\
      \sz \oo &  \sz \oo & \sz  1  \\
      \sz 1 &  \sz \oo &  \sz \oo
    \end{array} \right)
  \lambda^7
  =
  \left(
    \begin{array}{ccc}
      \sz \oo &  \sz \oo &  \sz \oo  \\
      \sz \oo &  \sz \oo &  \sz -i  \\
      \sz \oo &  \sz i &  \sz \oo
    \end{array} \right)        ,
\end{eqnarray}
and the $v^i, i=1\ldots 3$ are just three components vectors:
\begin{equation}
  v^1
  =
  \left(
    \begin{array}{c}
      -i\sqrt{2}  \\
      0\\
      0
    \end{array} \right)
  v^2
  =
  \left(
    \begin{array}{c}
      0\\
      i\sqrt{2}  \\
      0
    \end{array} \right)
  v^3
  =
  \left(
    \begin{array}{c}
      0\\
      0\\
      - i\sqrt{2}
    \end{array} \right).
\end{equation}
The generators of $G_2$ have been normalized in the usual way:
\begin{equation}
  {\rm Tr} \left(T^a {T^b}^\dagger\right) = \frac{1}{2} \delta^{ab}.
\end{equation}
Defining:
\begin{equation}
  S^i = T^{8+i} \ \ {\rm and} \ \ {\bar S}^i = T^{11+i}
  \ \ {\rm for} \  i=1\ldots 3,
\end{equation}
the $G_2$ algebra then reads:
\begin{eqnarray}
  &&
  [ T^a , T^b ] = \frac{i}{2\sqrt{2}} f^{ab}{}_c T^c ;
  \left[ {T^a} , S^i \right] = S^k {(T^a)}^{k i} ;
  \left[ T^a,{\bar S}^i \right] = - {(T^a)}^{i k} {\bar S}^k;\\
  &&
  \left[ S^i,S^j \right]= \frac{1}{\sqrt{3}} \epsilon^{ijk} {\bar S}^k;
  \left[ {\bar S}^i,{\bar S}^j \right]
  = -\frac{1}{\sqrt{3}} \epsilon^{ijk}  S^k;
  \left[ S^i,{\bar S}^j \right] = (T^a)^{ij} T^a .
\end{eqnarray}
where the $ f^{ab}{}_c$ are just the usual structure constants
associated to the Gell-Mann matrices and $\epsilon^{ijk}$ is the
totally antisymmetric tensor.  From the normalization factors
in~(\ref{eq:SU3inG2}), we get that the gauge coupling of the $SU(3)$
subgroup is relating in 6D to the gauge coupling of $G_2$ by
$g_{6D}^{SU(3)}=g_{6D}^{G_2}/\sqrt{2}$.  After compactification to 4D,
the gauge coupling of $SU(2)$ is given by
$g_{4D}^{SU(2)}=g_{6D}^{G_2}/(2\sqrt{2}\pi R)$ while the gauge
coupling of the $U(1)_Y$ normalized to $Y=(1/2,1/2,-1)$ in the
fundamental of $SU(3)$ is $g_{4D}^{Y}=g_{6D}^{G_2}/(2\sqrt{6}\pi R)$.
As announced in the introduction, we get $\sin^2 \theta_W=1/4$.

The components of the gauge fields are defined by
\begin{equation}
  A_M (x,y,z) = A^a_M (x,y,z) T^a ;
\end{equation}
and the orbifold conditions~(\ref{eq:OrbifoldOnGauge}) take the
form~(\ref{eq:AaM}) with the block diagonal matrices
\begin{eqnarray}
  &
  {\tilde R} = {\rm diag} \left(1,1,1,1,
    \left( \begin{array}{cc}
        &-1\\
        1
      \end{array}\right)
  \right)
  \vphantom{\begin{array}{c} 1\\1\\1\\1\end{array}},
  \\
  &
  {\tilde \Lambda}=  {\rm diag} \left(1,1,1,
    \left( \begin{array}{cc}
        &1\\
        -1
      \end{array}\right),
    \left( \begin{array}{cc}
        &1\\
        -1
      \end{array}\right),
    1,i,i,-1,-i,-i,-1
  \right).
\end{eqnarray}
%

%%%%%%%%%%%%%%%%%%%%%%%%%%%%%%%%%%%%%%%%%%%%
%%%%%%%%%%%%%%%%%%%%%%%%%%%%%%%%%%%%%%%%%%%%%%%%%%%%%
%%%%%%%%%%%%%%%%%%%%%%%%%%%%%%%%%%%%%%%%%%%%%%%%%%%%%
\section{The $G_2$ Group Element Implementing Parity for the $Z_4$ Orbifold}
\label{App:G2element}
\setcounter{equation}{0}
\setcounter{footnote}{0}
%%%%%%%%%%%%%%%%%%%%%%%%%%%%%%%%%%%%%%%%%%%%%%%%%%%%%
%%%%%%%%%%%%%%%%%%%%%%%%%%%%%%%%%%%%%%%%%%%%%%%%%%%%%

In this appendix we show, that it is possible to find a group element
$P$ in $G_2$ which satisfies $P U^{-1} = U P$.  The $Z_4$ orbifold is
acting on the fundamental of $G_2$ by the matrix $U= {\rm diag}(i, i,
-1, -i, -i, -1,1)$.  On can think of the interchange $(1,1)
\leftrightarrow (4,4)$, $(2,2)\leftrightarrow (5,6)$ by $P$ to convert
$U$ to $U^{-1}$, but actually such an element does not exist in $G_2$.
However, the interchange $(1,1) \leftrightarrow (5,5)$ and $(2,2)
\leftrightarrow (4,4)$ instead achieves $P^{-1} U P =U^{-1}$.

In order to show this, we are going to construct the matrix $P$ from
the generators of $G_2$ given above.  Let us look at the hermitian
combination
\begin{equation}
  R =  (S^3 + {\bar S}^3)
\end{equation}
It is straightforward to see that the Lie group element $g(\theta) =
e^{i \theta \sqrt{3} R}$ is given by
\begin{equation}
  g(\theta)
  = \left(
    \begin{array}{ccc|ccc|c}
      \cos\sfrac{\theta}{2} &  &  &  & -i \sin \sfrac{\theta}{2} &  & \\
      & \cos\sfrac{\theta}{2} &  & i \sin \sfrac{\theta}{2} &  &  &  \\
      \tvbas{8}
      &  & \sfrac{1+\cos\theta}{2}  &  &  &  \sfrac{1-\cos\theta}{2}
      & \sfrac{i}{\sqrt{2}} \sin\theta\\
      \hline
      \tv{15}
      & i \sin \sfrac{\theta}{2}&  &\cos\sfrac{\theta}{2} &  &  &  \\
      -i \sin\sfrac{\theta}{2}  &  &  &  & \cos\sfrac{\theta}{2}  &  &  \\
      \tvbas{8}
      &  & \sfrac{1-\cos\theta}{2} &  &  & \sfrac{1+\cos\theta}{2} 
      & \sfrac{-i}{\sqrt{2}} \sin\theta \\
      \hline
      \tv{15}
      &  & \sfrac{i}{\sqrt{2}} \sin \theta &  &  
      &  \sfrac{-i}{\sqrt{2}} \sin\theta  & \cos\theta
    \end{array} \right).
\end{equation}
Now taking $\theta = \pi$,
\begin{equation}
  P = g(\pi) = \left(
    \begin{array}{ccc|ccc|c}
      &  &  &  & -i &  &  \\
      &  &  & i &  &  &  \\
      &  &  &  &  & 1 &  \\
      \hline
      & i &  &  &  &  &  \\
      -i &  &  &  &  &  &  \\
      &  & 1 &  &  &  &  \\
      \hline
      &  &  &  &  &  & -1
    \end{array} \right).
\end{equation}
By construction $P$ is a group element of $G_2$ and one can easily
check that $P U P^{-1} = U^{-1}$, as desired.

%%%%%%%%%%%%%%%%%%%%%%%%%%%%%%%%%%%%%%%%%
%%%%%%%%%%%%%%%%%%%%%%%%%%%%%%%%%%%%%%%%%

\end{document}